\documentclass[a4paper,11pt]{article}
\pdfoutput=1
\usepackage[utf8x]{inputenc}
\usepackage{jheppub,color,subfigure,slashed,hyperref}
\usepackage{multirow}

\hypersetup{bookmarks=true,unicode=true,pdftoolbar=true,pdfmenubar=true,
	pdffitwindow=false,pdfstartview={FitH}, pdfsubject={},
	pdfcreator={},pdfproducer={}, pdfkeywords={},
	pdfnewwindow=true,colorlinks=true,
	linkcolor=blue,citecolor=magenta,filecolor=magenta,urlcolor=cyan}

\newcommand{\be}{\begin{equation}}
\newcommand{\ee}{\end{equation}}
\newcommand{\bea}{\begin{eqnarray}}
\newcommand{\eea}{\end{eqnarray}}
\newcommand{\eq}[1]{Eq.~(\ref{#1})}

\newcommand{\nn}{\nonumber}

\newcommand{\tev}{\,\text{TeV}}
\newcommand{\gev}{\,\text{GeV}}
\newcommand{\mev}{\,\text{MeV}}

\def\stw{s_{\theta_W}}
\def\ctw{c_{\theta_W}}
\def\ttw{t_{\theta_W}}
\def\lra#1{\overset{\text{\scriptsize$\leftrightarrow$}}{#1}}

\makeatletter
\def\section{\@startsection {section}{1}{\z@}{-3.5ex plus -1ex minus
 -.2ex}{2.3ex plus .2ex}{\large\bf}}
\def\subsection{\@startsection{subsection}{2}{\z@}{-3.25ex plus -1ex
minus -.2ex}{1.5ex plus .2ex}{\normalsize\bf}}
\makeatother
\makeatletter

\@addtoreset{equation}{section}

\makeatother

\newcommand{\madgraph}{{\sc MadGraph}}
\newcommand{\pythia}{{\sc Pythia 8}}

\newcommand{\fastjet}{{\sc FastJet}}
\newcommand{\ufo}{{\sc UFO}}
\newcommand{\feynrules}{{\sc FeynRules}}
\newcommand{\mcfm}{{\sc MCFM 7}}
\newcommand{\hepmc}{{\sc HepMC}}
\newcommand{\rivet}{{\sc Rivet}}
\newcommand{\powheg}{{\sc POWHEG BOX V2}}

\begin{document}

\begin{flushright} CERN-TH-2020-220, TUM-HEP-1309/20, IPPP/20/68 \end{flushright}

\title{A fully differential SMEFT analysis of the golden channel using the method of moments}
\author[a]{Shankha Banerjee,}
\author[b]{Rick~S.~Gupta,}
\author[b]{Oscar Ochoa-Valeriano,}
\author[b]{Michael~Spannowsky,}
\author[c]{Elena Venturini}
\affiliation[a]{CERN, Theoretical Physics Department, CH-1211 Geneva 23, Switzerland,}
\affiliation[b]{Institute for Particle Physics Phenomenology,\\Durham University, South Road, Durham, DH1 3LE, United Kingdom,}
\affiliation[c]{Technische Universit\"{a}t M\"{u}nchen, Physik-Department, \\James-Franck-Stra{\ss}e 1, 85748, Garching, Germany}

\emailAdd{shankha.banerjee@cern.ch}
\emailAdd{sandeepan.gupta@durham.ac.uk}
\emailAdd{oscar.ochoa-valeriano@durham.ac.uk}
\emailAdd{michael.spannowsky@durham.ac.uk}
\emailAdd{elena.venturini@tum.de}
\date{\today}

\abstract{The Method of Moments is a powerful framework to disentangle the relative contributions of amplitudes of a specific process to its various phase space regions. We apply this method to carry out a fully differential analysis of  the Higgs decay channel $h \to 4\ell$ and constrain gauge-Higgs coupling modifications parametrised by dimension-six effective operators. We find that this analysis approach provides very good constraints and minimises degeneracies in the parameter space of the effective theory. By combining the decay $h \to 4\ell$ with Higgs-associated production processes, $Wh$ and $Zh$, we obtain the strongest reported bounds on anomalous gauge-Higgs couplings.}

\maketitle

\section{Introduction}
\label{sec: intro}

One of the main goals of the Large Hadron Collider (LHC) is to understand the precise nature of electroweak symmetry breaking. The most direct way to probe this is via a measurements of the couplings of the Higgs boson to the weak bosons.  While the LHC measurements have already established that the Higgs boson couples to the gauge bosons~\cite{Aad:2020mkp, CMS-PAS-HIG-19-001, Aaboud:2018jqu, Aad:2019lpq, Sirunyan:2020tzo}, a full resolution of the tensor structure of these couplings is not possible without a thorough differential study of the relevant processes, namely-- Higgs decays to gauge bosons, Higgsstrahlung and Higgs production in vector boson fusion. New tensor structures for these couplings are unavoidable once we include operators from the next order in the Standard Model Effective Field Theory (SMEFT) expansion (see Ref.~\cite{Buchmuller:1985jz, Giudice:2007fh, Grzadkowski:2010es, Gupta:2011be,Gupta:2012mi,Banerjee:2012xc, Gupta:2012fy, Banerjee:2013apa, Gupta:2013zza,Elias-Miro:2013eta, Contino:2013kra, Falkowski:2014tna,Englert:2014cva, Gupta:2014rxa, Amar:2014fpa, Buschmann:2014sia, Craig:2014una, Ellis:2014dva, Ellis:2014jta, Banerjee:2015bla, Englert:2015hrx, Ghosh:2015gpa, Degrande:2016dqg,Cohen:2016bsd, Ge:2016zro, Contino:2016jqw, Biekotter:2016ecg, deBlas:2016ojx, Denizli:2017pyu, Azatov:2017kzw, Barklow:2017suo, Brivio:2017vri, Barklow:2017awn, Khanpour:2017cfq, Englert:2017aqb, panico,Franceschini:2017xkh, banerjee1, Grojean:2018dqj,Biekotter:2018rhp, Goncalves:2018ptp,Gomez-Ambrosio:2018pnl, Freitas:2019hbk,Banerjee:2019pks, Azatov:2019xxn, Banerjee:2019twi, Biekotter:2020flu, Rao:2020hel, Araz:2020zyh} for other relevant SMEFT studies). As we await the arrival of large volumes of data from the high luminosity runs of the LHC, the development of differential strategies to probe the gauge-Higgs couplings is thus of great relevance.

In this work we perform a fully differential analysis of the golden Higgs decay channel, $h \to 4 \ell$~\cite{Stolarski:2012ps, Chen:2014gka, Chen:2015iha, Chen:2016ofc, Gainer:2018qjm} using the `method of moments'~\cite{Dunietz:1990cj,Dighe:1998vk}. This technique would utilise the fact that in the SM, as well as the dimension-six (D6) SMEFT, the full angular distribution of the four leptons can be written as a sum of  a set of `basis functions'--a fact that can be elegantly understood to be a consequence of angular momentum conservation. An extraction of the coefficients of these nine functions, the so-called angular moments, from the observed data thus amounts to a study of the full multi-dimensional differential distribution of the four final-state leptons. The method of moments provides a well defined way to perform this extraction using a technique analogous to Fourier analysis.  Such a granular analysis is, in-fact, essential to disentangle the contribution of the different tensor structures of the gauge-Higgs coupling correcting this process.

The method of moments provides a transparent alternative to complement other multivariate techniques such as Optimal Observables~\cite{oo1,oo2},  the matrix element likelihood analysis (MELA)~\cite{MELA, Gritsan:2020pib} or other recently proposed methodologies involving Machine Learning~\cite{wulzernew}. In particular, MELA is currently the main technique used by the experiments to study the tensor structure of gauge-Higgs couplings.  While our method may not be able to surpass the matrix element method in power, we believe it is a way to achieve comparable bounds in a more physically transparent and intuitive way. This is because if an angular moment shows a deviation from its Standard Model (SM) value, it would be possible to pinpoint both the helicity amplitudes as well as  experimental distributions that are getting EFT contributions.

This work can be seen as a continuation of Ref.~\cite{Banerjee:2019twi} where the method of moments  was used to obtain the strongest reported projections for the measurement of the gauge-Higgs couplings in the Higgstrahlung processes, $pp \to Wh/Zh$. In this work we will finally combine the projections from the $h \to 4 \ell$ channel with the results from Ref.~\cite{Banerjee:2019twi} for the $pp \to Wh/Zh$ process. As we will see, these processes probe complimentary directions in the EFT space so that their combination results in highly stringent bounds on the EFT deformations of the Higgs coupling to gauge bosons.

The paper is divided as follows. In section~\ref{sec:sec2}, we list the relevant operators in the Warsaw basis that contribute to various vertex deformations for the $pp \to 4\ell$ process. We derive the angular dependence on the amplitude in section~\ref{sec:sec3}. The method of moments and its estimates along with the estimates of the uncertainties are described in section~\ref{sec:sec4}. In section~\ref{sec:sec6}, a detailed collider analysis including the angular extraction, is performed. We present our results in section~\ref{sec:sec7}. Finally, we conclude and present our outlook in section~\ref{sec:sec8}.

\section{The $pp \to h\to 4 \ell$ in the dimension-six SMEFT}
\label{sec:sec2}
Here we study the gluon initiated process $pp \to h \to  4 \ell$, where $\ell = e, \mu$. Given our analysis strategy, to be discussed in Sec.~\ref{sec:sec6},  the dominant contribution to this process in the SM  is from the  $h \to Z^{(*)}Z^* \to 4 \ell$ process where one of the $Z$-bosons is definitely off-shell~\footnote{In particular, we will consider the final states with $\ell = e, \mu$ with a hard cut on the missing transverse energy ($\slashed{E}_T$) and with no jets passing the trigger criteria. This mostly eliminates electrons and muons arising from $Z \to \tau \tau$ followed by leptonic $\tau$ decays.}. Furthermore, in a large majority of events we  find one of the $Z$ bosons to be  on-shell because of the resonant enhancement of the corresponding $Z$ propagator.

In the SMEFT the $h \to 4\ell$ decay can, in principle, arise also from topologies with anomalously large $h \bar{\ell} \ell$ couplings but such contributions will still not be comparable to the  $h \to Z^{(*)}Z^* \to 4 \ell$   contribution if we impose current bounds on the $h \bar{\ell} \ell$ couplings~\cite{Altmannshofer:2015qra, Aad:2020xfq}.  Another possibility is that the leptons arise from intermediate photons, however production of intermediate photons is loop suppressed and would require enhancement by at least an order of magnitude to have any impact; this would be easily ruled out by the bounds on the branching ratios for $h \to \gamma \gamma$~\cite{Sirunyan:2018ouh, CMS:2020omd} and $Z \gamma$~\cite{Aad:2020plj, Sirunyan:2018tbk} (see Ref.~\cite{HLLHC} for HL-LHC projections) and will not be considered further.  In addition  the contact interaction $hZ\bar{\ell}\ell$ gives a new diagram not present in the SM.  Finally, the SM diagram for this process gets EFT corrections at various vertices, \textit{i.e.}, $ggh$, $hZZ$, $Z\bar{\ell}\ell$. All these  corrections are summarised by the following Lagrangian written in the broken phase~\cite{Gupta:2014rxa, Pomarol:2014dya}~\footnote{We have ignored dipole structures for $Z\bar{\ell}\ell$ coupling deviations as contributions due these  to $Z\to \ell \ell$ are negligible because  there is no interference  with the SM  amplitude without  lepton mass insertions.},
\bea
\Delta {\cal L}_6 &\supset&  \delta \hat{g}^h_{ZZ} \, \frac{2 m_Z^2}{v}h \frac{Z^\mu Z_\mu}{2} + \sum_{\ell} \delta g^Z_{\ell} Z_\mu \bar{\ell} \gamma^\mu \ell +\sum_{\ell} g^h_{Z\ell} \, \frac{h}{v}Z_\mu \bar{\ell} \gamma^\mu \ell \nn \\ 
&+& \kappa_{ZZ} \, \frac{h}{2v} Z^{\mu\nu}Z_{\mu\nu}+ \tilde{\kappa}_{ZZ} \, \frac{h}{2v}Z^{\mu\nu}\tilde{Z}_{\mu\nu} 
\label{anam}
\eea
where, for brevity, we have just included the first generation leptons for the couplings with the $Z$-bosons, such that  $\ell=e_L, e_R, \nu^e_L$. The Lagrangian is assumed to be extended to the remaining two generations, such that the couplings $\delta g^Z_{\ell}$ and $g^{h}_{Z\ell}$ are flavour diagonal and universal in the interaction basis. This allows us to impose strong constraints on these couplings~\cite{Pomarol:2013zra, Falkowski:2014tna}. This assumption is theoretically well-motivated and can be obtained by including the leading terms after imposing Minimal Flavour Violation (MFV)~\cite{DAmbrosio:2002vsn}. In the above Lagrangian we have omitted any EFT corrections related to the production of the Higgs boson as all these corrections cannot be parametrised by local Lagrangian terms. We discuss these at the end of the section.

\begin{table}[t]
\small
\centering
\begin{tabular}{c}
\begin{tabular}{||c|c||}
\hline
&\\
${\cal O}_{H\square}=(H^\dagger H) \square (H^\dagger H)$ & ${\cal O}_{HB}= |H|^2 B_{\mu\nu}B^{\mu\nu}$ \\
\rule{0pt}{4ex} ${\cal O}_{HD}=(H^\dagger  {D}_\mu H)^*(H^\dagger  {D}_\mu H)$ & ${\cal O}_{HWB}=  H^\dagger \sigma^a H W^a_{\mu\nu}B^{\mu\nu}$ \\
\rule{0pt}{4ex} ${\cal O}_{H\ell}=i H^\dagger \lra{D}_\mu H \bar{e}_R  \gamma^\mu e_R$ & ${\cal O}_{H{W}}= |H|^2 W_{\mu\nu}{W}^{\mu\nu}$ \\
\rule{0pt}{4ex} ${\cal O}^{(1)}_{HL}=i H^\dagger  \lra{D}_\mu H \bar{L} \gamma^\mu L$ & ${\cal O}_{H\tilde{B}}= |H|^2 B_{\mu\nu}\tilde{B}^{\mu\nu}$ \\
\rule{0pt}{4ex} ${\cal O}^{(3)}_{HL}=i H^\dagger \sigma^a \lra{D}_\mu H \bar{L}  \sigma^a \gamma^\mu L$ & ${\cal O}_{H\tilde{W}B}=  H^\dagger \sigma^a H W^a_{\mu\nu}\tilde{B}^{\mu\nu}$ \\
\rule{0pt}{4ex} ${\cal O}_{HtG}=\bar{Q}_3\tilde{H}T^A\sigma_{\mu\nu}t_R G^{A \mu\nu}$ & ${\cal O}_{H\tilde{W}}= |H|^2 W^a_{\mu\nu}\tilde{W}^{a \mu\nu}$ \\
\rule{0pt}{4ex} ${\cal O}_{HbG}=\bar{Q}_3\tilde{H}T^A\sigma_{\mu\nu}b_R G^{A \mu\nu}$  & ${\cal O}_{y_b}=   |H|^2(\bar{Q}_3 H b_R+h.c).$ \\
\rule{0pt}{4ex} ${\cal O}_{HG}=(H^\dagger H)G^A_{\mu\nu}G^{A \mu\nu}$ & ${\cal O}_{y_t}=   |H|^2(\bar{Q}_3 H t_R+h.c).$ \\
&\\
\hline
 \end{tabular}
\end{tabular}
\caption{List of dimension-six operators in the Warsaw basis which contribute to the anomalous $hVV^*/hV\bar{f}f$, the effective Higgs-gluon, Yukawa and chromomagnetic couplings in \eq{anam}. Details about the notations can be found in Ref.~\cite{Grzadkowski:2010es}.}
\label{opers}
\end{table}
The parameterisation in the above Lagrangian also holds for a non-linearly realised electroweak symmetry~\cite{Isidori:2013cga} and in this scenario, all the above couplings must be considered as independent. On the other hand, if  electroweak symmetry is linearly realised, the aforementioned vertices, in the unitary gauge, arise from operators containing the Higgs doublet. The list of  operators in the Warsaw basis~\cite{Grzadkowski:2010es} that contribute to this process, including those that affect the Higgs boson production are  shown in Table~\ref{opers}; these contribute to the said vertices as follows,
 \bea
 \label{wilson}
   \delta g^Z_{\ell}&=&-\frac{g Y_{\ell} \stw}{\ctw^2}\frac{v^2}{\Lambda^2} c_{HWB} -\frac{g}{\ctw}\frac{v^2}{\Lambda^2}(|T_3^{\ell}|c^{(1)}_{HL}-T_3^{\ell} c^{(3)}_{HL}+(1/2-|T_3^{\ell}|)c_{H\ell})\nn\\&+&\frac{\delta m^2_Z}{m^2_Z}\frac{g}{2\ctw\stw^2}(T_3 \ctw^2+Y_{\ell} \stw^2)\nn\\
  \delta \hat{g}^h_{ZZ}&=&\frac{v^2}{\Lambda^2}\left(c_{H\square}+\frac{c_{HD}}{4}\right)\nn\\
  g^h_{Z\ell}&=&- \frac{2 g}{\ctw}\frac{v^2}{\Lambda^2}(|T_3^{\ell}|c^{(1)}_{HL}-T_3^{\ell} c^{(3)}_{HL}+(1/2-|T_3^{\ell}|)c_{H\ell})\nn\\
  \kappa_{ZZ}&=&\frac{2 v^2}{ \Lambda^2}(\ctw^2 c_{HW}+\stw^2 c_{HB}+ \stw \ctw c_{HWB})\nn\\
   \kappa_{GG}&=&\frac{2 v^2}{ \Lambda^2}c_{HG}\nn\\
  \tilde{\kappa}_{ZZ}&=&\frac{2 v^2}{ \Lambda^2}(\ctw^2 c_{H\tilde{W}}+\stw^2 c_{H\tilde{B}}+ \stw \ctw c_{H\tilde{W}B}),
 \eea
 where, $(m_W, m_Z,\alpha_{em})$ are our input parameters. In the equation for $\delta g^Z_{\ell}$ above, the term
\bea
\frac{\delta m^2_Z}{m^2_Z}= \frac{v^2}{\Lambda^2}(2 \ttw c_{HWB}+\frac{c_{HD}}{2}),
\eea
explicitly shows the contribution of two of the aforementioned operators to the shift in $m_Z$; one of the input parameters.

  
Of all the anomalous couplings in Eq.~\ref{anam}, only $\delta \hat{g}^h_{ZZ}, \kappa_{ZZ}$ and $\tilde{\kappa}_{ZZ}$ would be eventually relevant for us. This is because the other couplings can be measured or constrained much more stringently in other processes. First,  note that LEP1~\cite{ALEPH:2005ab} has put per-mille level constraints on $\delta g^Z_{\ell}$ from the partial $Z$-decay measurements of $\Gamma(Z \to \ell\bar{\ell})$ parameters~\cite{Pomarol:2013zra, Falkowski:2014tna}; the corrections due to these couplings would thus be neglected.
 
 
As far as the corrections due to the couplings, $g^h_{Z\ell}$, are concerned they can be ignored as these couplings can be very stringently constrained at HL-LHC using the D6 SMEFT correlations. This is because these couplings receive contributions from the same operators as $\delta g^Z_{\ell}$, apart from $c_{WB}$ and $c_{HD}$ that only contribute to $\delta g^Z_{\ell}$.  The Wilson coefficients $c_{WB}$ and $c_{HD}$ can be constrained by their contribution to the anomalous charged Triple Gauge Couplings (TGCs)~\cite{Hagiwara:1986vm},
\bea
\delta g^Z_{1}&=&\frac{1}{2 \stw^2}\frac{\delta m_Z^2}{m_Z^2} \nn \\
\delta\kappa_{\gamma}&=&\frac{1}{\ttw} \frac{v^2}{\Lambda^2} c_{HWB}\;.
\eea
Using the expressions for $g^h_{Z\ell}$, $\delta g^Z_{\ell}$ in Eq.~\ref{wilson} and the above expressions for the TGCs we obtain the following relationship,
\begin{equation}
    g^h_{Z\ell}= \frac{2 g}{c_{\theta_W}}Y_{\ell} t_{\theta_W}^2 \delta \kappa_\gamma+2 \delta g^Z_{\ell}- \frac{2 g}{c_{\theta_W}}(T^{\ell}_3 c_{\theta_W}^2 + Y_{\ell} s_{\theta_W}^2)\delta g_1^Z
\label{contact}    
\end{equation}
derived also in~\cite{Gupta:2014rxa, Pomarol:2014dya}. Whereas, the $\delta g^Z_{\ell}$ couplings are very stringently constrained by $Z$-pole measurements as discussed above, per-mille level bounds are also expected for the TGCs at the HL-LHC~ \cite{wulzer, montull}. The contact term couplings, $g^h_{Z\ell}$ can, therefore, be tightly constrained at the HL-LHC using the correlation in Eq.~\ref{contact}. While we will consider the effect of the $g^h_{Z\ell}$ couplings in our theoretical discussions, these couplings will be eventually neglected in our final numerical analysis that would lead to bounds only at the  percent level~\footnote{The four-point contact interactions, $g^h_{Z\ell}$  can also be directly constrained at future $e^+e^-$ colliders, running at TeV scale energies. It was shown in  Ref.~\cite{Banerjee:2018bio},  that the analogous contact term couplings involving quarks can be constrained at the per-mille level or stronger at the HL-LHC; one can thus expect bounds at a similar level for $g^h_{Z\ell}$ as at lepton colliders.}.

We have still not considered corrections to Higgs production in the $pp \to h \to Z^{(*)}Z^* \to 4 \ell$ process. These involve five other operators, namely the Yukawa operators, ${\cal O}_{y_b, y_t}$, the chromomagnetic operator,  ${\cal O}_{Htg, Hbg}$ , and the Higgs gluon effective operator, ${\cal O}_{HG}$ as discussed in Refs.~\cite{Degrande:2012gr, Deutschmann:2017qum}. In this work, we perform our analysis in the Higgs rest-frame and hence the effects of these operators only appear as an overall factor that affects the total rate but can be decoupled as far as the lepton distributions are concerned. These effects thus effectively redefine $\delta \hat{g}^{h}_{ZZ}$, which also affects only the rate and has no differential signature, 

\begin{equation}
\label{eq:Higgsprod}
    (1+\delta \hat{g}^{h}_{ZZ}) \rightarrow (1+\delta \hat{g}^h_{ZZ})(1  + f(c_{HG}, c_{yb}, c_{yt}, c_{HtG}, c_{HbG})),
\end{equation}
where  $f(c_{HG}, c_{yb}, c_{yt},  c_{HtG}, c_{HbG})$, a linear combination of the aforementioned Wilson coefficients, can be obtained from the results of  Refs.~\cite{Degrande:2012gr, Deutschmann:2017qum}. In order to constrain and disentangle these EFT corrections in the production sector, it is necessary to study the production of $h+$ jet and $t\bar{t}h$, in conjunction and include other Higgs decay channels.

As far as EFT contributions to the non-Higgs background are concerned, the main corrections to the dominant background from the  $q\bar{q} \to 4\ell$  process come from the TGCs and the $Z$-coupling deviations, $\delta g^Z_\ell$, both of which would be strongly constrained in other processes as discussed above. As the $gg \to 4\ell$ background is much smaller (about a percent, see Sec.~\ref{sec:sec6}) we will not consider EFT modifications, for instance via anomalous $\bar{t}tg$ and $\bar{t}tZ$ vertices. Finally, one should also notice that a deviation from SM in the $gg\to h\to 4\ell$ amplitude modifies its interference with the $gg\to ZZ\to 4\ell$ continuum as well. However, we will neglect in our analysis this contribution to the total cross-section since at invariant masses around $ m_h$  this effect is negligible with respect to the Higgs production channel~\cite{rikkert}.

\section{Angular dependence of the $h \to 4l$ amplitude}
\label{sec:sec3}


As we discussed in the previous section, the predominant contribution to the  $pp\to h\to 4\ell$ process is from  $gg\to h \to Z^{(*)}Z^*\to 4\ell$ in the SM. In SMEFT at dimension-six level there can be two different kinds of processes that dominantly contribute to $h\to 4\ell$. The first one is mediated, as in SM, by two intermediate $Z$ bosons, taking into account the $\delta \hat{g}^h_{ZZ}$, $\kappa_{ZZ}$ and $\tilde{\kappa}_{ZZ}$ (see Eq.~\ref{contact}) modifications of the $hZZ$ vertex. The second one corresponds to an amplitude containing an effective $hZ\bar{\ell}\ell$ contact interaction (induced by the ${\cal O}_{HL,\ell}$ SMEFT operators, see Sec.~\ref{sec:sec2}), followed by the decay of the single $Z$ that is produced. In both cases, the $4\ell$ final state is made of two fermion currents, at least one of which is emitted via vector decay. Each current contains an outgoing fermion and an outgoing anti-fermion having opposite helicities, in the massless limit; thus, in the following, each one of these 2-fermion states will be denote as an $\ell_+ \ell_-$ system.

\begin{figure}[t]
    \centering
    \includegraphics[width=1 \textwidth]{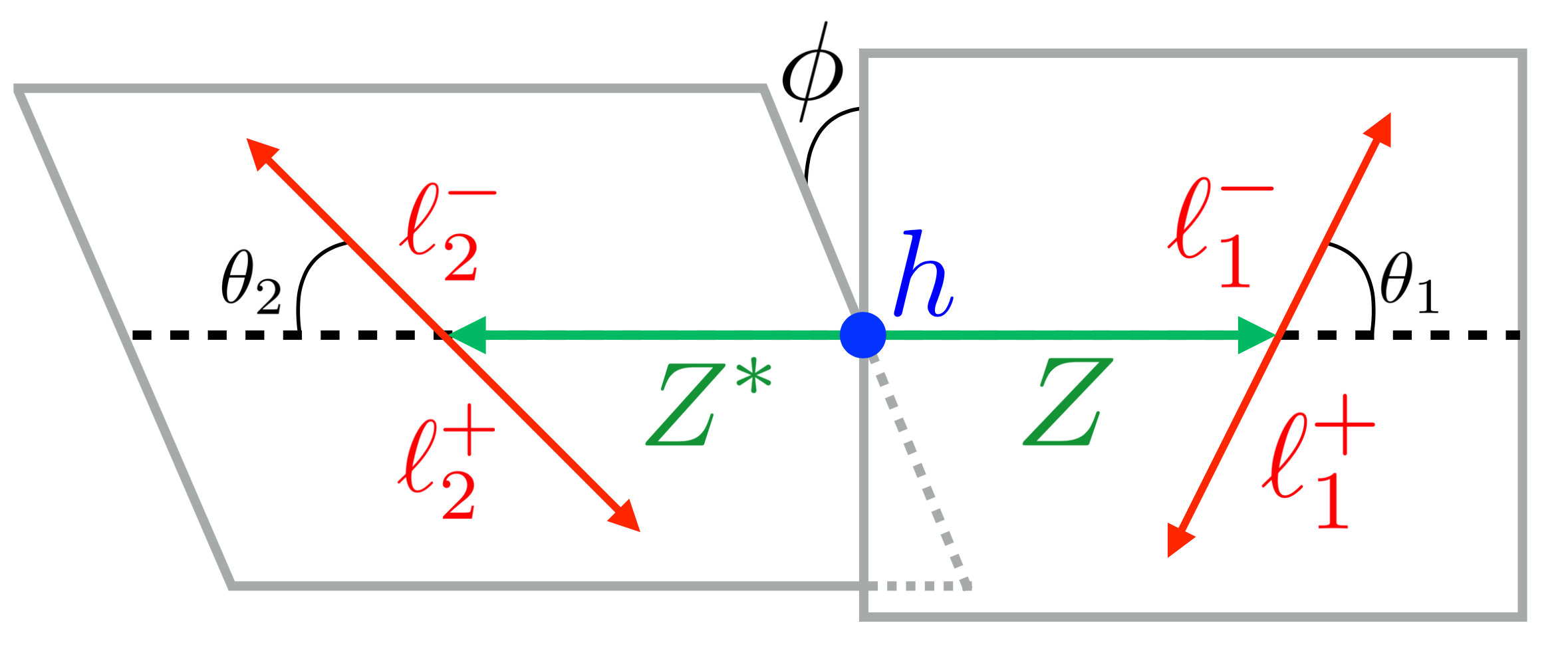}
    \caption{Diagram showing the definition of the different scattering angles for the sequential decay $h \to ZZ^* \to (\ell_1^- \ell_1^+)(\ell_2^- \ell_2^+)$. Note that two different reference frames are used to compute i)~the azimuthal angle $\phi$ between the planes formed from the lepton pairs in the Higgs rest frame, and ii)~the polar angle $\theta_i \, (i = 1, 2)$ of the lepton in the rest frame of its parent $Z$ boson. Each frame is characterised by the presence of back-to-back objects. We adopt the convention used in ref.~\cite{Godbole_2007}.}
    \label{fig:scatt_angles}
\end{figure}

The goal here is the analysis of the differential distribution with respect to the angular variables described in the following. The three angles required to define the final state fermions are shown in Fig.~\ref{fig:scatt_angles}, where the angle definitions assume that the leptons have a fixed charge. Theoretically, however, it is much more convenient to express the angular distribution for final-state fermions with definite helicity.  For the purposes of the subsections, Sec.~\ref{subsec:sec31} and Sec.~\ref{subsec:sec32}, the reader should assume that these angles refer to the positive helicity lepton and not the negatively charged lepton.  We will provide the translation to the experimentally accessible distribution with respect to final state leptons with a fixed charge later in Sec.~\ref{subsec:sec33}.   The angles $\theta_1$ and $\theta_2$ are the polar angles that the momentum of a lepton $\ell_{+}$, with chosen positive helicity, forms with the direction of motion of the center of mass of its parent $\ell_+ \ell_-$ system. This is evaluated   in the $\ell_+ \ell_-$ center of mass frame where the fermion and the antifermion are back to back. For example, when $\ell_+$ and $\ell_-$ come from a $Z$ decay, $\theta_i$ is measured in the $Z$ rest frame, where $\theta_i=0$ corresponds to the direction of motion of the parent $Z$ in the Higgs rest frame, where $i=1,2$ refers to the two $\ell_+ \ell_-$ system. Furthermore, we consider, for each $\ell_+ \ell_-$ system, the azimuthal angle $\varphi_i$ that describes the orientation of the plane individuated by the two fermion momenta, evaluated in the Higgs (or equivalently $4\ell$) center of mass frame; we, then, define $\phi$ as the relative azimuthal angle between the two planes. For more details on  definition of these angles see Sec.~\ref{sec:sec6}.

As we have commented above, the structure of the interactions is such that each $\ell_+ \ell_-$ system has two opposite helicity fermions with $\lambda=\pm 1/2$. Then, the  angular dependence of the amplitude  is determined uniquely by the angular momentum quantum numbers of both the 2-fermion states, in the $\ell_+ \ell_-$ center of mass frame, namely by the total angular momentum $J$ of each 2-fermion state and by the projection $M$ along the $\ell_+ \ell_-$ direction of motion. More specifically, the total amplitude is proportional to the Wigner functions $d^J_{M,\Delta\lambda =1}(\theta_i,\varphi_i)$ for both the $\ell_+ \ell_-$ final states, where the angles are evaluated considering the momentum of the positive helicity fermions and $\Delta \lambda=1$ is the helicity difference between $\ell_+$ and $\ell_-$.

When $\ell_+ \ell_-$ come from a decay of a single particle, $J$ and $M$ are given respectively by the total spin and helicity of the intermediate unstable state. In our case, the decaying particle is a $Z$ boson, implying $J=1$ and $M=0,\pm 1$ according to the helicity of the vector. Therefore, the angular dependence will be described by the following Wigner functions:
\bea
&&d^1_{+ 1,\Delta \lambda=1}(\theta_i,\varphi_i)=\cos^2(\theta_i/2)e^{+ i \varphi_i}\, , \quad d^1_{- 1,\Delta \lambda=1}(\theta_i,\varphi_i)=\sin^2(\theta_i/2)e^{- i \varphi_i}\nonumber\\&& d^1_{0,\Delta\lambda =1}(\theta_i,\varphi_i)=\frac{\sin\theta_i}{\sqrt{2}}\, .
\eea
As a consequence, the study of $h\to ZZ^*$ and $h\to Z\ell\ell$ helicity amplitudes is crucial and it will be done in the next subsections. Even in the case of amplitude with insertion of a $hZ\bar{\ell}\ell$ contact term, the angular modulation turns out to be completely described by $d^1_{M,\Delta \lambda=1}(\theta,\varphi)$ functions, as discussed in details in Sec.~\ref{subsec:sec32}. 

\subsection{Helicity amplitudes for $ZZ^*$ production}
\label{subsec:sec31}

We consider a scalar particle which decays into 2 states; in the scalar rest frame, the system has zero total angular momentum. Therefore, in this frame the two final objects, with opposite momenta, must have same helicities, in order to guarantee $M=0$ for the full final state. It is the case of $h\to ZZ^*$ in the Higgs rest frame: the helicity configurations allowed for the final states are $Z_+Z_+$, $Z_-Z_-$ and $Z_0Z_0$. We call the corresponding helicity amplitudes as $A_{++}$, $A_{--}$ and $A_{00}$ .

The possible deviations from SM in the $hZZ$ interactions are induced by the two CP-even effective couplings $\delta \hat{g}^h_{ZZ}$ and $\kappa_{ZZ}$, that contribute to  all the three helicity amplitudes, and the CP-odd coupling $\tilde{\kappa}_{ZZ}$ that contributes only to the transverse amplitudes. The three $h\to ZZ^*$ amplitudes are:
\begin{align}
A_{++} &= -2\,  \frac{(\delta \hat{g}^h_{ZZ}+1)\,  m_Z^2}{v}+2\, \frac{\kappa_{ZZ}}{v} \,  \gamma_a \, m_Z \, {m_Z*} -2 i \, \frac{\tilde{\kappa}_{ZZ}}{v} \, \gamma_b \, m_Z \, {m_Z*}\label{eq:App}\\
A_{--} &= -2\,  \frac{(\delta \hat{g}^h_{ZZ}+1)\,  m_Z^2}{v}+2\, \frac{\kappa_{ZZ}}{v} \,  \gamma_a \, m_Z \, {m_Z*} +2 i \, \frac{\tilde{\kappa}_{ZZ}}{v} \, \gamma_b \, m_Z \, {m_Z*}\\
A_{00} &=  -2\,  \frac{(\delta \hat{g}^h_{ZZ}+1)\,  m_z^2}{v} \, \gamma_a - \,2 \, \frac{\kappa_{ZZ}}{v} \,\frac{1}{m_Z \, m_{Z^*}} \\
\nonumber
\end{align}
with
\begin{align}
\gamma_a &=  \, \frac{1}{m_Z \, {m_Z*} } \, (E_1 E_2 +|\vec{q}|^2) =  \frac{1}{m_Z \, {m_Z*} } \, q_Z \cdot q_{Z^*} \label{eq:gammaa}\\
\gamma_b & = \, \frac{1}{m_Z \, {m_Z*} }  \, |\vec{q}| (E_1+E_2) = \, \frac{1}{m_Z \, {m_Z*} }  \, |\vec{q}| m_h \label{eq:gammab}\\
\nonumber
\end{align} 
where $\vec{q}=\vec{q_Z}=-\vec{q_{Z^*}}$ are the 3-momenta in the Higgs rest frame, $m_Z^2=q_Z^2$ and $m_{Z^*}^2=q_{Z^*}^2$ are the squared invariant masses of the on-shell and off-shell $Z$ respectively. Above, the term independent from effective couplings $\delta \hat{g}^h_{ZZ}$, ${\kappa}_{ZZ}$ and $\tilde{\kappa}_{ZZ}$ describes the SM amplitude.

We have analysed above the $h\to ZZ^*$ helicity amplitudes; however, as previously discussed, the helicities of the vector bosons determine the dependence of the amplitude  on the final lepton angles. In fact, a $h\to ZZ^*\to \ell_+\ell_-\ell_+\ell_-$ helicity amplitude, which is obtained summing over the polarisations of the intermediate vector bosons, is given by:
\begin{align}
\label{eq:ampl_dec}
&{\cal M}(h\to ZZ^*\to \ell^1_+\ell^1_-\ell^2_+\ell^2_-)=g^Z_{\ell_1} g^{Z^*}_{\ell_2} A(h\to ZZ^*\to \ell^1_+\ell^1_-\ell^2_+\ell^2_-)\sim \\
& \sum_{\bar{\lambda}\bar{\lambda}^\prime}A(h\to Z_{\bar{\lambda}} Z^*_{\bar{\lambda}'})\frac{-g^Z_{\ell_1}}{q_{Z}^2-m_Z^2+i \Gamma_Z m_Z}  A(Z_{\bar{\lambda}}\to \ell_+^1\ell_-^1) \frac{-g^{Z^*}_{\ell_2}}{q_{Z^*}^2-m_Z^2+i \Gamma_Z m_Z} A(Z^*_{\bar{\lambda}'}\to \ell_+^2\ell_-^2)\\
&\propto \sum_{\bar{\lambda}} A(h\to Z_{\bar{\lambda}} Z^*_{\bar{\lambda}})\frac{-g^Z_{\ell_1}}{q_{Z}^2-m_Z^2+i \Gamma_Z m_Z}  d^1_{\bar{\lambda},\Delta \lambda=1}(\theta_1,\varphi_1)\frac{-g^{Z^*}_{\ell_2}}{q_{Z^*}^2-m_Z^2+i \Gamma_Z m_Z}d^1_{\bar{\lambda},\Delta \lambda=1}(\theta_2,-\varphi_2)\label{eq:ampl_Wig}\\ \, ,
\nonumber
\end{align}
where $A(h\to ZZ^*\to \ell^1_+\ell^1_-\ell^2_+\ell^2_-)$ and $A(Z^{(*)}_{\bar{\lambda}^{(')}}\to \ell_+^{1(2)}\ell_-^{1(2)})$ are the amplitudes with the $Z$-lepton couplings factorised in front. 

The parameter $g^{Z, Z^*}_{l_i}= \frac{g}{2\ctw}(T_{3\ell}- Q_{\ell}\stw^2)$ is the SM coupling of the $Z$ boson  which depends on the chirality of the fermions, $\ell= \ell_{L,R}$. For a fixed helicity the chirality is different for fermion and anti fermion fields: positive helicity corresponds to LH chirality in the case of positively charged leptons and to RH chirality in the case of negatively charged leptons.  A more detailed discussion about relation between helicity and charge (and chirality) will be presented in Sec.~\ref{subsec:sec33}. In Eq.~\ref{eq:ampl_dec},  $\bar{\lambda}$ and $\bar{\lambda'}$ are the helicities of $Z$ and $Z^*$ and we have used in the second line the fact they must be equal in the Higgs rest frame. The angles are defined, for the two lepton pairs, as explained above and the minus for $\varphi_2$ arises due to the choice of the same reference frame for both the azimuthal angles as shown in Sec.~\ref{sec:sec6}. We see explicitly that the form of the angular dependence, encoded by the Wigner functions, is determined for each term in the sum by the polarisation of intermediate vectors. The production helicity amplitude $A(h\to Z_{\bar{\lambda}} Z^*_{\bar{\lambda}})$ is different for different polarisations $\bar{\lambda}$ and receives BSM corrections (it is the only one under the assumptions applied here). Therefore, deviations from SM modify the differential distribution in the lepton angular variables. On the other hand, the Breit Wigner propagators do not depend on the helicity of the vector and thus can be factored out for all the terms with different angular modulations.
Thus, the $A_{++}$, $A_{--}$ and $A_{00}$ amplitudes are respectively multiplied by the following angular functions, that are products of two $d^1_{\bar{\lambda},\Delta \lambda=1}$:
\begin{align}
f_{++} & = \cos^2(\theta_1/2)\cos^2(\theta_2/2) e^{+i \phi}\label{eq:fpp}\\
f_{--} & = \sin^2(\theta_1/2)\sin^2(\theta_2/2) e^{-i \phi}\\
f_{00} & = \sin(\theta_1) \sin(\theta_2)\label{eq:f00}\, .
\end{align}
where $\phi=\varphi_1-\varphi_2$.

In the discussion above, we have assumed that one of the two $Z$ bosons is produced on-shell, but both $Z$'s could be off-shell, in which case the $m_Z m_{Z^*}$ terms in Eq.~\ref{eq:App}-\ref{eq:gammab} become $m_{Z_1^*} m_{Z_2^*}$. However, the cross-section of a process with one on-shell $Z$ is enhanced, as one can also see from Eq.~\ref{eq:ampl_dec}; as a consequence, in the majority of the cases one pair of leptons will have invariant mass around the $m_Z$ mass (see Sec.~\ref{sec:sec6} for a more detailed discussion). We should notice that one could consider the angles for the negative helicity fermion, in which case the $d^1_{\bar{\lambda},\Delta \lambda=-1}$ Wigner functions enter in the amplitude.


\subsection{Helicity amplitudes for $ Z\ell\ell$ production through $hZ\bar{\ell}\ell$ contact interactions}
\label{subsec:sec32}

We now consider the $g_{Z\ell}^h$ contribution to  the angular distributions derived in Sec.~\ref{subsec:sec31}. Although we do not consider the effect of these interactions in our numerical analysis--as they can be probed more precisely in other processes as discussed in Sec.~\ref{sec:sec2}-- we  will keep them in all our analytical expressions. 

We will show here how the $g_{Z\ell}^h$ contribution to the full $h\to 4\ell$ amplitude can be simply seen as a shift in the coupling of the leptons to the off-shell Z, $g^{Z^*}_{\ell_2}$,   in Eq.~\ref{eq:ampl_dec}. In the considered $h Z_\mu \bar{\ell}\gamma^\mu \ell$ operator, the fermion bilinear transforms as a vector under Lorentz transformations; in other words, the contact interaction has the form $hZ_\mu V^\mu$, where $V^\mu$ is a spin 1 object. Therefore, the ${\ell}_+\ell_-$ system involved in the $g_{Z\ell}^h$ vertex has intrinsic angular momentum $J=1$. Furthermore, as previously explained, independently from the form of the interaction, in a $h\to ZV$ decay evaluated in the Higgs rest frame, $Z$ and $V$ must have same helicity. As a consequence, the $M$ angular momentum quantum number for the ${\ell}_+\ell_-$ system, in its center of mass frame, is fixed to be equal to the helicity of the emitted $Z$. Therefore, the dependence on lepton angular variables is given by $d^1_{M,\Delta \lambda=1}(\theta,\phi)$ and is determined by the helicity $\bar{\lambda}=M$ of the $Z$ boson involved in the contact interaction; it is the same angular modulation that we would have if the leptons were coming from the decay of an additional intermediate $Z$. Furthermore,  as the $hZ_\mu V^\mu$ form of the contact term is  analogous to the SM  $h Z_\mu Z^\mu$ vertex, we obtain the same amplitude for all the three helicity configurations. Thus, the $g_{Z\ell}^h$ contribution to $h\to Z2\ell\to 4\ell$ corresponds to a shift in the SM ${\cal M}(h \to ZZ^*\to 4\ell)$ amplitude in Eq.~\ref{eq:ampl_dec} that can be express as a shift in the $g^{Z^{*}}_{\ell_2}$ coupling. Explicit computation confirms this heuristic reasoning and we obtain for the shift:
\be
\label{eq:gZfh}
g^{Z^{*}}_{\ell_2}\to g^{Z^{*}}_{\ell_2}-g_{Z \ell_2}^h \, \frac{m_Z^2-m_{Z^*}^2-i\Gamma_Z\, m_Z}{2m_Z^2} \, .
\ee

So far, we have assumed that the $Z$ is produced on-shell, but one could have a $g_{Z\ell}^h$ contact interaction with emission of an off-shell $Z$ as well. However, it will be suppressed with respect to the on-shell case, due to the absence of resonance enhancement.

\bigskip


\subsection{Visible angular modulation}
\label{subsec:sec33}

In the previous sections, we have considered $h\to 4\ell$ helicity amplitudes, evaluating the angles for leptons with $\lambda=+\frac{1}{2}$. However, one cannot experimentally have access to the helicities of the final state leptons. For this reason, in the angular analysis, the electric charge is fixed instead; in particular, in our analysis we consider angles that describe the emission direction of negatively charged fermions (see Fig.~\ref{fig:scatt_angles} and Sec.~\ref{sec:sec6} for details). Then, in the physical $h\to 4\ell$ process that can be studied at colliders, each final lepton has definite charge but not helicity. The squared amplitude, if expressed in terms of helicity amplitudes, should thus be summed over the four possible helicity configurations $\bar{\ell}^1_{\pm} {\ell}^1_{\mp}, \, \bar{\ell}^2_{\pm} {\ell}^2_{\mp}$, where $\bar{\ell}$ and $\ell$ are respectively positively and negatively charged leptons, namely anti-fermions and fermions, and $\pm$ stand for the helicities (recall that these are equal and opposite within each fermion pair):
\be
|{\cal M}(h\to \bar{\ell}^1\ell^1\bar{\ell}^2\ell^2)|^2= \sum_{\lambda ,\lambda'} |{\cal M}(h\to \bar{\ell}^1_{-\lambda} \ell^1_\lambda\bar{\ell}^2_{-\lambda'}\ell^2_{\lambda'})|^2
\ee
The differential distribution is computed with respect to the measured $\ell^i$ emission angles, that coincide with the $\theta^i$ and $\varphi^i$ entering in Eq.~\ref{eq:ampl_Wig} in the cases in which $\ell^i_\lambda$ has positive helicity $\lambda=+\frac{1}{2}$. On the other hand, in the terms where the negatively charged fermion is $\ell^i_{-}$ the measured angles $\theta^i$ and $\varphi^i$ refer to the negative helicity leptons, which implies that the angular distribution is described by the Wigner functions $d^1_{\bar{\lambda},\Delta \lambda=-1}(\theta,\varphi)=d^1_{\bar{\lambda},\Delta \lambda=1}(\pi-\theta,\pi+\varphi)$.
Then, remembering that positive helicity corresponds to positive charge in the case of LH fermions and to negative charge for RH fermions, if we take into account the $\theta$ and $\phi$ (or equivalently $\varphi$) variables for negatively charged leptons, the angular definition will be left unchanged for RH leptons, while in case of LH chirality in the angular functions $f_{++}$, $f_{--}$, $f_{00}$ of Eqs.~\ref{eq:fpp}-~\ref{eq:f00} we should apply the substitution $(\theta,\phi)\to (\pi - \theta ,\phi +\pi)$. Thus, different helicity $h\to 4\ell$ amplitudes correspond to different chirality configurations for the final fermions and therefore are associated to different $Z$-lepton couplings. Then, the total squared amplitude of the observed process can be expressed as:
\begin{align}
|{\cal M}(h\to \bar{\ell}^1\ell^1\bar{\ell}^2\ell^2)|^2&= \Bigg( {g^Z_{l_R}}^2{g^{Z^*}_{l_R}}^2|A(\theta_1,\theta_2,\phi)|^2+{g^Z_{l_L}}^2{g^{Z^*}_{l_L}}^2 |A(\pi-\theta_1,\pi-\theta_2,\phi)|^2+ \nn \\
& + {g^Z_{l_L}}^2{g^{Z^*}_{l_R}}^2|A(\pi-\theta_1,\theta_2,\pi+\phi)|^2+{g^Z_{l_R}}^2{g^{Z^*}_{l_L}}^2|A(\theta_1,\pi-\theta_2,\pi+\phi)|^2\Big)
\label{eq:sum_hel}
\end{align}
and $A(\theta_1,\theta_2,\phi)=A(h\to \ell^1_+\ell^1_-\ell^2_+\ell^2_-)(\theta_1,\theta_2,\phi)$ is the helicity amplitude in Eq.~\ref{eq:ampl_dec} in which the angles are evaluated for positive helicity fermions and where the $Z$-lepton couplings $g^{Z^{(*)}}_{l_{L(R)}}$ have been factorised out. 



\section{The Method of Moments}
\label{sec:sec4}

In this section we use the results of the previous section to  show that the $h \to 4\ell$ squared amplitude can be written as a sum of a set of angular functions both in the SM and D6 SMEFT.  We will  describe the method for extraction of the coefficients of these  functions, the so-called angular moments, and discuss the associated uncertainty estimates.

\subsection{Angular moments for $h \to 4\ell$}
\label{angmom}
The cross-section of the process $gg\to h\to 4\ell$, induced by the two contributions studied in Secs.~\ref{subsec:sec31} and~\ref{subsec:sec32}, is obtained summing the squared helicity amplitudes as in Eq.~\ref{eq:sum_hel}. Therefore, as one can see by considering Eqs.~\ref{eq:fpp}-~\ref{eq:f00}, it is a linear combination of the following 9 functions of the final lepton angles:

\bea
f_1 & = &\sin^2(\theta_1) \sin^2(\theta_2) \nn \\
f_2 & = & (\cos^2(\theta_1)+1)( \cos^2(\theta_2)+1) \nn \\
f_3 & = & \sin(2\theta_1) \sin(2\theta_2) \cos(\phi) \nn \\
f_4 & = & (\cos^2(\theta_1)-1)( \cos^2(\theta_2)-1)\cos(2\phi) \nn \\
f_5 & = & \sin(\theta_1)\sin(\theta_2)\cos(\phi) \nn \\
f_6 & = & \cos(\theta_1)\cos(\theta_2) \nn \\
f_7 & = & (\cos^2(\theta_1)-1)( \cos^2(\theta_2)-1)\sin(2\phi) \nn \\
f_8 & = & \sin(\theta_1)\sin(\theta_2)\sin(\phi)\nn \\
f_9 & = & \sin(2\theta_1) \sin(2\theta_2) \sin(\phi),
\label{funcs}
\eea
where the last 3 functions appear in the CP-odd terms, linear in the $\tilde{\kappa}_{ZZ}$ coupling and the angles are defined for negatively charged leptons as in Fig.~\ref{fig:scatt_angles}. Our results agree with Ref.~\cite{Godbole_2007}.

The angular moments are the coefficients of the 9 angular functions above, in the differential cross-section evaluated in the Higgs rest frame. As a function of the $ \delta \hat{g}^h_{ZZ}$, $\kappa_{ZZ}$, $ \tilde\kappa_{ZZ}$ and $g_{Zf}^h$ coefficients they are:

\bea
a_1 & = & {\cal G}^4\left((1+\delta a)+\frac{ b m_{Z^*} \gamma_b^2}{ m_Z\gamma_a}\right)^2 \nn \\
a_2 & = & {\cal G}^4\left(\frac{(1+\delta a)^2}{2\gamma_a^2}+\frac{2 c^2  m_{Z^*}^2 \gamma_b^2}{ m_Z^2 \gamma_a^2}\right) \nn \\
a_3 & = & -{\cal G}^4\left(\frac{1+\delta a}{2\gamma_a}+\frac{ b  m_{Z^*} \gamma_b^2}{2  m_Z \gamma_a}\right)^2 \nn \\
a_4 & = &{\cal G}^4\left( \frac{(1+\delta a)^2}{2\gamma_a^2}-\frac{2 c^2  m_{Z^*}^2 \gamma_b^2}{  m_Z^2\gamma_a^2}\right) \nn \\
a_5 & = & -\epsilon^2 {\cal G}^4 \left( \frac{2 (1+\delta a)^2}{\gamma_a} + \frac{2 (1+\delta a) b  m_{Z^*} \gamma_b^2}{ m_Z \gamma_a^2}\right) \nn \\
a_6 & = & \epsilon^2 {\cal G}^4 \left( \frac{2 (1+\delta a)^2}{\gamma_a^2} +\frac{8 c^2  m_{Z^*}^2 \gamma_b^2}{ m_Z^2 \gamma_a^2}\right) \nn \\
a_7 & = & {\cal G}^4 \frac{2 (1+\delta a) c  m_{Z^*} \gamma_b}{ m_Z \gamma_a^2} \nn \\
a_8 & = & - \epsilon^2 {\cal G}^4 \left( \frac{4 (1+\delta a) c  m_{Z^*} \gamma_b}{ m_Z \gamma_a}+ \frac{4 b c  m_{Z^*}^2 \gamma_b^3}{ m_Z^2 \gamma_a^2}\right) \nn \\
a_9 & = & {\cal G}^4\left(\frac{ (1+\delta a) c  m_{Z^*} \gamma_b}{m_Z \gamma_a}+ \frac{b c  m_{Z^*}^2 \gamma_b^3}{m_Z^2 \gamma_a^2}\right),
\eea
where

\bea
{\cal G}^4&=&((g^Z_{l_L})^2+(g^Z_{l_R})^2)((g^{Z^*}_{l_L})^2+(g^{Z^*}_{l_R})^2) \nn \\
\epsilon^2{\cal G}^4&=& ((g^Z_{l_L})^2-(g^Z_{l_R})^2)((g^{Z^*}_{l_L})^2-(g^{Z^*}_{l_R})^2),
\eea
contain the effect of the contact terms, $g^h_{Z\ell}$ via Eq.~\ref{eq:gZfh} and we have chosen a normalisation such that  $a_1^{\rm SM}={\cal G}^4$. Here, $\gamma_a$ and $\gamma_b$ are defined in Eqs.~\ref{eq:gammaa} and~\ref{eq:gammab} and
\begin{align}
\delta a &=   \delta \hat{g}^h_{ZZ} -\kappa_{ZZ} \gamma_a \frac{m_{Z^*}}{m_Z} \, \frac{m_Z^2-m_{Z^*}^2}{2m_Z^2} \nn \\
b&=\kappa_{ZZ} \nn \\
c&=-\frac{\tilde{\kappa}_{ZZ}}{2}
\end{align}

The angular moments, $a_5, a_6$ and $a_8$, are numerically suppressed in the SM as well as the EFT interference term. This is because, once the $g^h_{Z\ell}$ are stringently constrained by other processes as discussed in Sec.~\ref{sec:sec2}, the SM value  
\bea
\epsilon^2= \left(\frac{(g^Z_{l_L})^2-(g^Z_{l_R})^2}{(g^Z_{l_L})^2+(g^Z_{l_R})^2}\right)^2= 0.16^2= 0.026\, 
\eea
acts as a suppression factor. Thus, these moments  contribute only marginally as far as the final numerical bounds are concerned.

To understand this suppression better, consider for example how the $\cos(\theta_1)\cos(\theta_2)$ dependence arises in $f_6$ in Eq.~\ref{funcs}. The $\cos(\theta_1)\cos(\theta_2)$ term in the first and second terms in  Eq.~\ref{eq:sum_hel} does not change sign, but the substitution  $\cos(\theta_1)\cos(\theta_2)\to -\cos(\theta_1)\cos(\theta_2)$ is applied in the third and fourth terms where one of  $\ell^1$ or $\ell^2$ is a LH and negative helicity fermion. This gives a factor of $\epsilon^2$  in the total squared amplitude, making this contribution numerically small. In fact, there are cases, like this one, in which the helicity-charge interplay leads to a partial or almost complete cancellation of the angular differential distributions that we would have observed if we could have access to the helicities of the final state leptons. In general, by averaging over all the helicity configurations, we loose part of the information contained a priori in the angular modulations of Eqs.~\ref{eq:fpp}-~\ref{eq:f00}.

\subsection{The basic idea behind the method of moments}
\label{basic}
As seen in Sec.~\ref{angmom}, the squared amplitudes for our present process, can be written as a set of angular structures, $f_i(\theta_1,\theta_2,\phi)$, which are parameterised by the corresponding coefficients, the angular moments, $a_i$. In this section, we explain how to extract these coefficients by taking the best-possible advantage of all the available angular information. Even though a full likelihood fit can be appropriate, here we consider the method of moments~\cite{Dunietz:1990cj, james, Beaujean:2015xea}. This method is transparent and advantageous, especially when the number of events is not very large~\cite{Beaujean:2015xea}. For this method, an analog of Fourier analysis is used in order to extract the angular moments. Essentially, we seek weight functions, $w_i(\theta_1,\theta_2,\phi)$, that can extract all the coefficients, $a_i$ uniquely, i.e.,
\begin{align}
&&\int_0^\pi d\theta_1 \int_0^\pi d\theta_2 \int_0^{2\pi} d\phi \sum_i (a_i f_i) w_j\sin\theta_1 \sin\theta_2 = a_j, \nn \\
&\Rightarrow& \int_0^\pi d\theta_1 \int_0^\pi d\theta_2 \int_0^{2\pi} d\phi f_iw_j\sin\theta_1 \sin\theta_2 =\delta_{ij}.
\label{mom_def}
\end{align}
Upon assuming that these weight functions are linear combinations of the functions in the original basis, we can write
\begin{equation}
w_i = \lambda_{ij}f_j.
\end{equation}
We can use \eq{mom_def} to show that $\lambda_{ij}=M_{ij}^{-1}$, with,
\begin{equation}
\label{matrixM}
M_{ij}= \int_0^\pi d\theta_1 \int_0^\pi d\theta_2 \int_0^{2\pi} d\phi f_if_j\sin\theta_1 \sin\theta_2.
\end{equation}
For the set of basis functions listed in \eq{funcs}, the corresponding matrix is given by,
\begin{equation}
M = 
\left(
\begin{array}{ccccccccc}
\frac{512 \pi }{225} & \frac{128 \pi }{25} & 0 & 0 & 0 & 0 & 0 & 0 & 0 \\
\frac{128 \pi }{25} & \frac{6272 \pi }{225} & 0 & 0 & 0 & 0 & 0 & 0 & 0 \\
0 & 0 & \frac{256 \pi }{225} & 0 & 0 & 0 & 0 & 0 & 0 \\
0 & 0 & 0 & \frac{256 \pi }{225} & 0 & 0 & 0 & 0 & 0 \\
0 & 0 & 0 & 0 & \frac{16 \pi }{9} & 0 & 0 & 0 & 0 \\
0 & 0 & 0 & 0 & 0 & \frac{8 \pi }{9} & 0 & 0 & 0 \\
0 & 0 & 0 & 0 & 0 & 0 & \frac{256 \pi }{225} & 0 & 0 \\
0 & 0 & 0 & 0 & 0 & 0 & 0 & \frac{16 \pi }{9} & 0 \\
0 & 0 & 0 & 0 & 0 & 0 & 0 & 0 & \frac{256 \pi }{225} \\
\end{array}
\right)\,,
\end{equation}
where we organise the basis functions according to their order in \eq{funcs}.

It is advantageous to translate to a basis such that $M_{ij}$, and thus correspondingly its inverse $\lambda_{ij}$, are diagonal. We can achieve this by the following orthogonal rotation,
\begin{align}
\hat{f}_1&=\cos \beta f_{1}-\sin \beta f_{2},\nonumber\\
\hat{f}_2&=\sin \beta f_{1}+\cos \beta f_{2},
\end{align}
by an angle, 
\begin{equation}
\tan\beta = -\dfrac{1}{2}(5+\sqrt{29}). 
\end{equation}

In the fully-orthogonal basis, we have, $\vec{\hat{f}}=\{\hat{f}_1,\hat{f}_2, f_{3},f_{4}, f_{5},f_{6},f_{7},f_{8},f_{9}\}$. The rotated matrix $M\to \hat{M}$, thus reads,
\begin{equation}
\hat{M}=\hat{\lambda}_{ij}^{-1}=\mathrm{diag}\left(
\frac{64 \pi }{225} \xi_+,
\frac{64 \pi }{225}\xi_-,
\frac{256 \pi }{225},
\frac{256 \pi }{225},
\frac{16 \pi }{9},
\frac{8 \pi }{9},
\frac{256 \pi }{225},
\frac{16 \pi }{9},
\frac{256 \pi }{225}
\right)
\end{equation}
with $\xi_\pm = (53\pm9\sqrt{29})$. Thus, the weight functions in the rotated basis can be expressed as
\begin{equation}
\label{wi}
w_i = \hat{M}^{-1}_{ij}f_j.
\end{equation}
We can now convolute the various distributions from our events with these weight functions and extract the coefficients in the new diagonalised basis,
\bea
\label{basis1}
\{\hat{a}_{1},\hat{a}_{2},a_{3}, a_{4}, a_{5},a_{6}, a_{7}, a_{8}, a_{9}\}.
\eea
These coefficients can be rotated back in case we are seeking the moments in the original basis.

\subsection{Moments estimates and estimation of uncertainties}\label{sec:sec43}

In the last sections we have defined the angular moments $a_i$ and their extraction through weight functions defined over a continuous phase space; we want now to show how to estimate them starting from the observed experimental dataset, or in order to obtain projections, from Monte Carlo samples. We will also discuss how  to estimate the uncertainty in this procedure. 

One can notice, from Eq.~\ref{mom_def}, that the angular moments indicated there are the expectation values of the weights $w_j$ for a probability distribution $\sum_i a_i f_i$ normalised to the squared amplitude. Changing the normalisation to the total number $N$ of observed  events, we can consider the random variable
\be
\tilde{a}_i= N w_i \, .
\ee
for each event in the experimental dataset. Averaging over all events  the observed value for the angular moments can be obtained, 
\be
\label{basicEq}
{a}_i= N \bar{w}_i=\sum_{n=1}^{\hat{N}} w_i(\theta_{1,n},\theta_{2,n},\phi_n).
\ee
which is indeed the discretised version of Eq.~\ref{mom_def}. This is the procedure that must be used by the experiments to extract the angular moments.

  In the absence of the true experimental dataset we have used, for our projections,  separate Monte Carlo samples both with and without the EFT couplings turned on. The SM sample includes both Higgs and non-Higgs backgrounds. These Monte Carlo samples have a much larger number of events, $N^{SM, EFT}_{MC}$,  than the  number of events expected at 3 ab$^{-1}$, that we denote as $\hat{N}^{SM, EFT}$. As we will discuss in Sec.~\ref{sec:sec6}, for our statistical analysis to estimate the final bounds, we will take the SM expectation  to be the null-hypothesis and assume that the experiments observe an excess over the SM because of the presence of EFT terms.


The weight functions $w_i$, for a sufficiently large number of events, converge to a multivariate Gaussian distribution for which the  estimates of the expectation values and of the covariance matrix are given by
\begin{align}
\bar{w_i}&=\frac{1}{N_{MC}}\sum_{n=1}^{N_{MC}} w_i(\theta_{1,n},\theta_{2,n},\phi_n)\,,\label{eq:estM}\\
\sigma_{ij}&=\frac{1}{N_{MC}-1}\sum_{n=1}^{N_{MC}}\left[w_i-\bar{w_i}\right] \left[{w_j}-\bar{w_j}\right],
\end{align}
 both for the SM and EFT samples. We find that if we keep increasing $N^{SM, EFT}_{MC}$, as soon as it is large enough (order 100), the $\bar{w_i}$ and $\sigma_{ij}$ approach fixed values, since the estimates converge to the true values of the parameters. Our estimates for the $a_i$ expectation values for the expected (SM) and observed (EFT) events are then evaluated using
\be
{a}^{SM, EFT}_i=\langle {N}^{SM, EFT}\rangle\bar{w_i}^{SM, EFT}= \hat{N}^{SM, EFT}\bar{w_i}^{SM, EFT}.
\ee
In the above equation ${N}^{SM, EFT}$ is assumed to be a variable following the Poisson distribution with both mean and variance given by $\hat{N}^{SM, EFT}$. 

Note that the  $\bar{w_i}^{SM, EFT}$  is supposed to be evaluated in an expected/observed dataset with a number of events equal to  $\hat{N}^{SM, EFT}$, which is much smaller than the number of Monte-Carlo events. The $\bar{w_i}^{SM, EFT}$ in Eq.~\ref{eq:estM}, which gives a very good estimate of the $w_i$ expectation value, is a random variable whose covariance is given by
 \be
 \bar{\sigma}_{ij}=\frac{{\sigma}_{ij}}{\hat{N}}
 \ee
 both for the SM and EFT cases. The $\bar{\sigma}_{ij}$ values decrease for increasing $\hat{N}$, which is related to the fact that the $\bar{w}_i$ estimates become more precise for larger number of events.


 The statistical uncertainties of the estimated mean values are computed as covariances of functions of the random variables $N$  and $\bar{w}_i$:
\begin{align}
\label{eq:ai_errstat}
{\rm cov}(a_i,a_j)&=\sum_{kl} \bar{\sigma}_{kl} \left(\frac{\partial a_i}{\partial w_k}\frac{\partial a_j}{\partial w_l}\right)\Big|_{w=\bar{w},N=\hat{N}}
+\hat{N} \left(\frac{\partial a_i}{\partial N}\frac{\partial a_j}{\partial N}\right)\Big|_{w=\bar{w},N=\hat{N}}\\&=\left(\frac{\sqrt{\hat{N}}}{\hat{N}}\right)^2 a_i a_j +\hat{N} \sigma_{ij}.
\end{align}
As we take  SM to be our null-hypothesis, the uncertainties that will enter our final $\chi^2$ function in Sec.~\ref{sec:sec6} are the ones related to the SM expectation. We also take into account a flat systematic covariance on the SM prediction given by $\kappa_{\rm syst}^2 a^{\rm SM}_ia^{\rm SM}_j$ where we will take $\kappa_{\rm syst}=0.02$  following Ref.~\cite{ATLAS:2018jlh}~\footnote{One can deduce a total fractional systematic error of $0.039=\sqrt{0.035^2+0.016^2+0.006^2}$,  from Table 2 of Ref.~\cite{ATLAS:2018jlh} where the three values-- respectively corresponding to the experimental error, the theory error for the SM Higgs process and the theory error for the non-Higgs background-- have been added in quadrature. Note, however, that this is normalised with respect to the SM $gg\to h\to ZZ^*$ process whereas $\kappa_{\rm syst}$ above is normalised with respect to the full SM cross-section including non-Higgs backgrounds. Our value $\kappa_{\rm syst}=0.02$ is obtained using $\kappa_{\rm syst}~\hat{N}_{SM}=0.039~\hat{N}^h_{SM}$, where $\hat{N}^h_{SM}$ is the number of SM Higgs events. Note that, while  Ref.~\cite{ATLAS:2018jlh}  expresses this systematic error as a fraction of the SM Higgs rate, it actually includes, as a separate contribution, the error from the predominantly $q\bar{q}$ initiated non-Higgs background.}. $a_i^{SM}$ includes all SM contributions, including the Higgs process. Then, the total covariance of the set of estimated SM angular moments is
\begin{align}
\label{eq:ai_err}
\Sigma_{ij}= \left(\left(\frac{\sqrt{\hat{N}_{SM}}}{\hat{N}_{SM}}\right)^2+\kappa_{\rm syst}^2\right) a^{\rm SM}_i a^{\rm SM}_j +\hat{N}_{SM} \sigma^{\rm SM}_{ij}.
\end{align}


\section{Collider simulation}
\label{sec:sec6}

\subsection{Monte Carlo samples and analysis setup}
\label{sec:MCAS}

In this section, we discuss the collider analysis that helps us in obtaining bounds on the relevant operator combinations. We implement our \ufo~\cite{Degrande_2012} model with the help of \feynrules~\cite{Alloul_2014}. This model is required to generate the signal samples, including the interference and the squared terms ensuing from the dimension-six interactions.

We consider the $gg \to h \to 4\ell$ process, where the irreducible backgrounds are composed of the quark-initiated $q\bar{q} \to 4\ell$ and gluon-initiated $gg \to 4\ell$ processes, with $\ell = \{e, \mu, \tau\}$. Reducible backgrounds arise from processes where jets can be misidentified as charged leptons in the fiducial region of the detector; these are dominated by $Z/\gamma^* + \mathrm{jets}$, which we generate as $p p \to \ell^+ \ell^- + 2$ jets. Negligible sources of fake backgrounds include $t\bar{t}$, $WW + \mathrm{jets}$, and $WZ + \mathrm{jets}$~\footnote{We take into account those processes that yield exactly four parton-level visible objects (charged~leptons~+~jets) in the final state, where we consider channels with up to three jets. We find that the only non-negligible contribution arises from $Z/\gamma^* + \mathrm{jets}$.}.

We generate Monte Carlo events considering a centre-of-mass energy of $\sqrt{s}=14$~\tev. The SM- and EFT-driven $gg \to h \to 4\ell$ samples, as well as the reducible background $Z/\gamma^* + \mathrm{jets}$, are generated at leading order (LO) with \madgraph~\cite{Alwall_2014}, including the full decay chain. The quark-initiated $q\bar{q} \to 4\ell$ background samples are generated at next-to-leading order (NLO) with \powheg~\cite{Nason_2004, Frixione_2007, Alioli_2010}. The  \texttt{NNPDF31$\_$lo$\_$as$\_$0130} and \texttt{NNPDF31$\_$nlo$\_$hessian$\_$pdfs} \cite{thennpdfcollaboration2017parton} PDF sets are used to generate \madgraph \,samples and $q\bar{q} \to 4\ell$ events, respectively. The gluon-initiated $gg \to 4\ell$ background samples are generated at LO with \mcfm~\cite{Campbell_2010} using the \texttt{CTEQ6L}~\cite{Pumplin_2002} PDF set. All events are further passed on to \pythia~\cite{Sj_strand_2015} for parton shower and hadronisation. For the quark-initiated background events, we apply a generator-level invariant-mass cut for each pair of opposite-sign same-flavour (OSSF) leptons of $M_{\ell^+, \ell^-} \geq 4$~\gev. For the remaining samples we require $\left| \eta^\ell \right| \leq \mathrm{3}$, as well as $\Delta R(\ell_i, \ell_j) \geq \mathrm{0.015}$, where $\Delta R=\sqrt{\Delta \eta^2 + \Delta \phi^2}$ is the separation in the $\eta - \phi$ plane. In \madgraph, we impose an additional set of cuts, namely $p_T^{\ell_1} \geq \mathrm{15}$~\gev, $p_T^{\ell_{2,3}} \geq \mathrm{8}$~\gev \; \textrm{and} \; $p_T^{\ell_4} \geq \mathrm{3}$~\gev. For the $Z/\gamma^* + \mathrm{jets}$ samples we further apply $p_T^j > \mathrm{20}\gev$, $\left| y^j \right| \leq \mathrm{3}$, $\Delta R(j, \ell) \geq \mathrm{0.015}$, $\Delta R(j_m, j_n) \geq \mathrm{0.015}$, and $M_{2\ell, 2j} \in \mathrm{[95, 155]} \gev$. In the case of events generated using \mcfm, we require $p_T^\ell \geq \mathrm{3}$~\gev, $M_{\ell^+,\ell^-} \geq \mathrm{2.5}$~\gev, as well as $M_{4\ell} \geq \mathrm{70}$~\gev. Here, the indices on the leptons indicate their $p_T$ ordering, with $\ell_1$ being the hardest lepton.

Following the recommendations of the LHC Higgs cross-section working group~\cite{deFlorian:2227475} (LHC HXSWG, CERN Report 4), we scale the LO $gg \to h$ production cross-section of the Monte Carlo sample to the $\mathrm{N^3LO}$-accurate prediction for $M_h = 125$~\gev, obtaining an overall $K$-factor of 3.155, which we apply to the SM-, as well as to the EFT-driven samples~\footnote{Within our simulation framework we further set the width of the Higgs boson $\Gamma_h = 4.088$~\mev, for consistency with the LHC HXSWG.}. We assume a flat NNLO/NLO $K$-factor of 1.1 for the $q\bar{q} \to 4\ell$ background, given the differential cross-section for the quark-initiated process shown in Ref.~\cite{Grazzini_2015}. For the NNLO/LO scaling of the $gg \to 4\ell$ samples, we apply a flat $K$-factor of 2.27, as considered in the experimental search described in Ref.~\cite{Sirunyan_2017}. We further adopt a conservative approach by using a flat $K$-factor of 0.91 \cite{Campbell_2002} for the $Z/\gamma^* + \mathrm{jets}$ events. After reweighting, the signal-to-irreducible background ratio $S/B_{\mathrm{irr}}$ is found to be 0.00734. Here, by signal we mean the SM production of $gg \to h \to 4\ell$.

We base our analysis strategy on the experimental search~\footnote{We validate our analysis against the experimental search, and find that our results are compatible with the experimental numbers within 96\%.} described in Ref. \cite{CMS-PAS-HIG-19-001}. A simplified detector analysis is performed on the stable final-state particles using \hepmc~\cite{Dobbs:2001ck} and \fastjet~\cite{Cacciari_2012}. Visible objects are selected if they fulfill $|\eta| < 4.7$ and $p_T > 0.5$~\gev. Electrons (muons) are preselected within the geometrical acceptance $|\eta| < 2.5 \, (2.4)$, with $p_T > 7 \, (5)$~\gev, and are in turn isolated by demanding that the total hadronic activity around a cone radius of $R = 0.3$ centred in the lepton's direction must be less than $35\%$ of its $p_T$. The overall missing transverse momentum $\slashed{E}_T$ is the magnitude of the total transverse momentum calculated from all preselected particles, and its direction is opposite to these transverse momenta. Jets are clustered using the anti-$k_t$ algorithm~\cite{Cacciari_2008} with a radius parameter 0.4 and $p_T > 30$~\gev. We simulate the detector response in the reconstruction of electrons, muons, and jets, by applying a Gaussian smearing~\cite{Buckley_2020}, as implemented in \rivet~\cite{BUCKLEY20132803}, to their energy, $p_T$, and 3-momentum components, respectively. For leptons, these mass- and direction-preserving smearing functions are applied before selection, whereas for jets the mass-preserving smearing is applied after clustering. We assume a flat leptonic reconstruction efficiency of 0.92 and consider a rapidity-dependent jet-to-electron fake rate of 0.016 (0.044) for jets with $\left| y^j \right| < \mathrm{1.48}$ $\left(1.48 < \left| y^j \right| < \mathrm{2.5}\right)$~\cite{cmscollaboration2020electron}.

\begin{figure}[t]
    \centering
    \includegraphics[width=1.0 \textwidth]{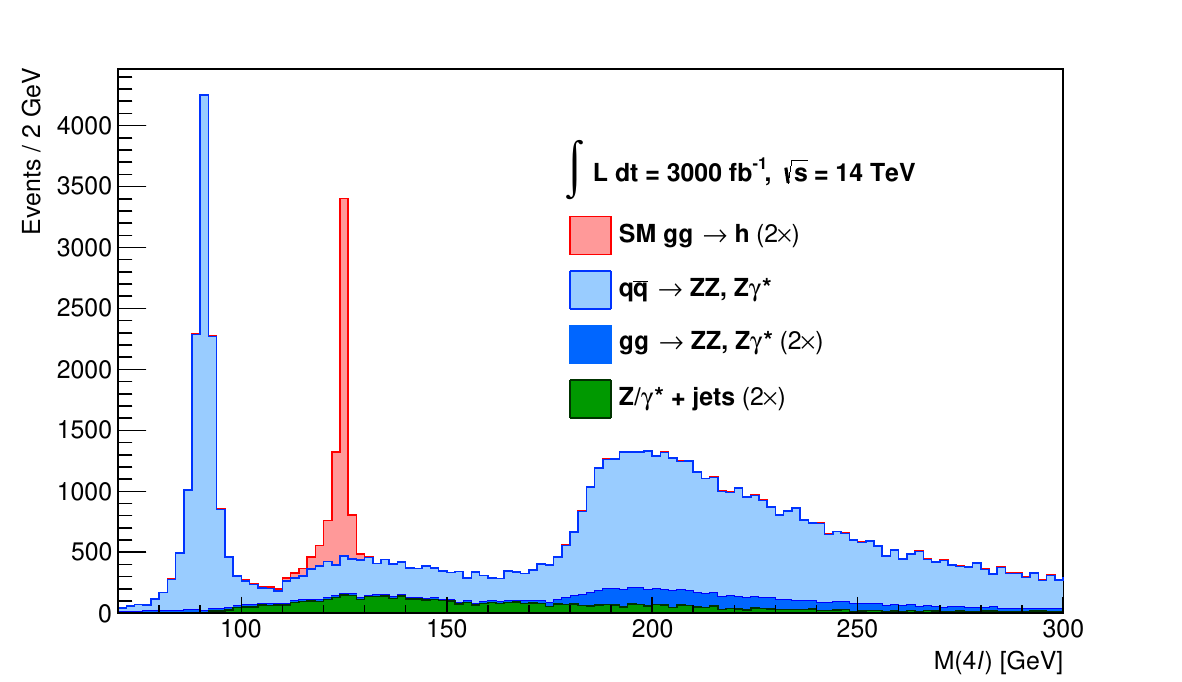}
    \caption{Invariant mass distribution $M(4\ell)$ of the 4-lepton system after reconstruction of $Z_1 Z_2$ pairs. Stacked histograms for SM $gg \to h$ (red) signal, quark-initiated $q\bar{q} \to 4\ell$ (light blue), gluon-initiated $gg \to 4\ell$ (dark blue), and $Z/\gamma^* + \mathrm{jets}$ (green) backgrounds, are normalised to the expected number of events at the HL-LHC. Except for the quark-initiated background, distributions are scaled ($2\times$) for visualisation purposes.}
    \label{fig:m4l}
\end{figure}

\begin{table}[t]
\begin{center}
\small
\begin{tabular}{|c|c|c|c|}\hline

\multirow{1}{*}{Selection cut} & 
\multirow{1}{*}{SM $gg \to h$} & 
\multirow{1}{*}{$q\bar{q} \to 4\ell$} & 
\multirow{1}{*}{$gg \to 4\ell$} \\ \hline \hline \hline

\multirow{1}{*}{Jet veto} & 
\multirow{1}{*}{0.419} & 
\multirow{1}{*}{0.779} & 
\multirow{1}{*}{0.319} \\ \hline

\multirow{1}{*}{$\slashed{E}_T < 25$~\gev} & 
\multirow{1}{*}{0.348} &
\multirow{1}{*}{0.667} & 
\multirow{1}{*}{0.248} \\ \hline

\multirow{1}{*}{2 pairs of isolated OSSF leptons,} & \multirow{1}{*}{} & \multirow{1}{*}{} & \multirow{1}{*}{} \\
\multirow{1}{*}{$\Delta R(\ell_i, \ell_j) > 0.02$,} &
\multirow{1}{*}{0.127} &
\multirow{1}{*}{0.036} &
\multirow{1}{*}{0.130} \\
\multirow{1}{*}{$M_{\ell^+, \ell^{\prime-}} > 4$~\gev} & \multirow{1}{*}{} & \multirow{1}{*}{} & \multirow{1}{*}{} \\ \hline

\multirow{1}{*}{$p_{T,\ell_1} > 20 \, \mathrm{GeV}$, $p_{T,\ell_2} > 10 \, \mathrm{GeV}$, $p_{T,\ell_3} > 10 \, \mathrm{GeV}$} &
\multirow{1}{*}{0.121} &
\multirow{1}{*}{0.031} &
\multirow{1}{*}{0.124} \\ \hline

\multirow{1}{*}{$M(Z_1) \in [40, 120]$~\gev, $M(Z_2) \in [12, 120]$~\gev} &
\multirow{1}{*}{0.110} &
\multirow{1}{*}{0.021} &
\multirow{1}{*}{0.112} \\ \hline

\multirow{1}{*}{$M(4\ell) \in [118, 130]$~\gev} &
\multirow{1}{*}{0.095} &
\multirow{1}{*}{0.001} &
\multirow{1}{*}{0.001} \\ \hline

\end{tabular}
\caption{Set of cuts showing the impact of each stage of the selection on the fraction of retained Monte Carlo events for the SM-driven $gg \to h \to 4\ell$ process, as well as on the $q\bar{q} \to 4\ell$ and $gg \to 4\ell$ irreducible backgrounds.}
\label{table:cutflow}
\end{center}
\end{table}

\begin{figure}
	\begin{center}
		\includegraphics[scale=.375]{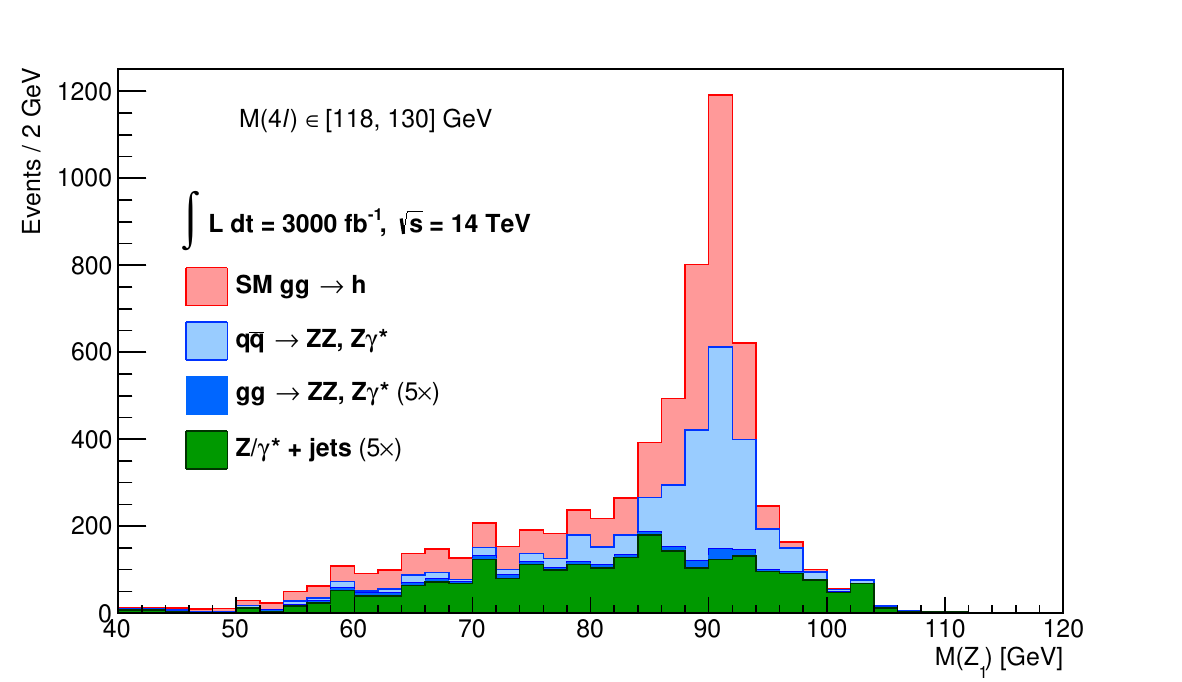}  
		\includegraphics[scale=.375]{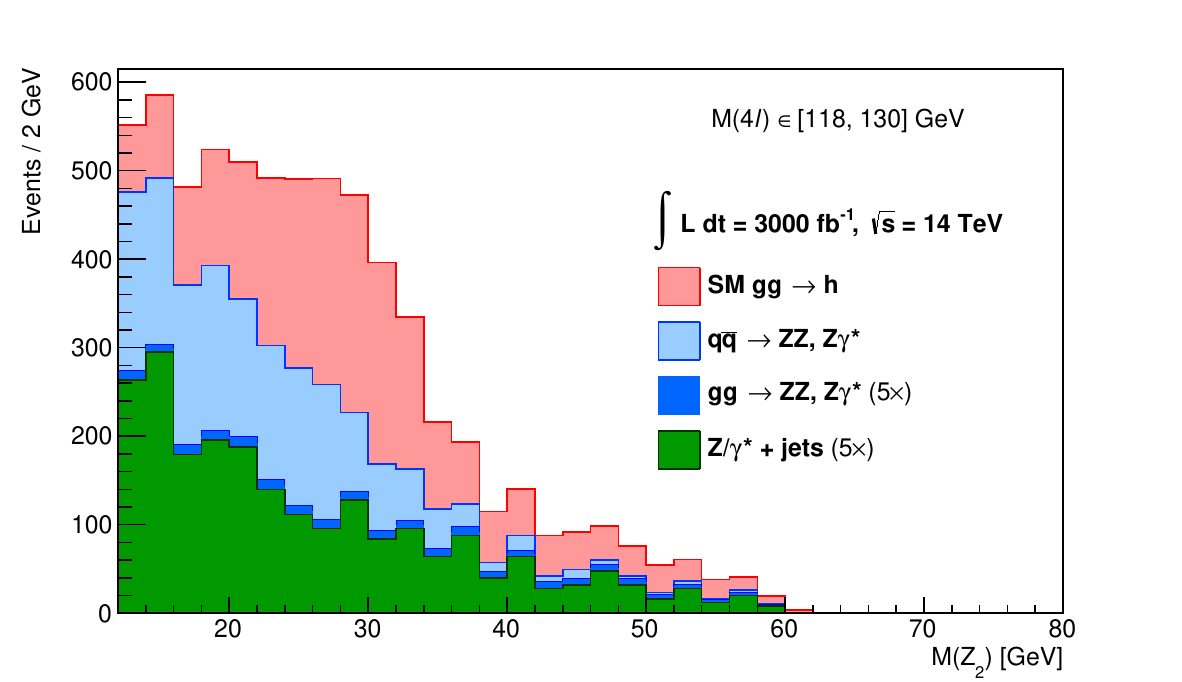} 
	\end{center}
	\caption{Invariant mass distribution $M(Z_i)$ of the (left) $Z_1$ and (right) $Z_2$ candidates after defining the signal region $M(4\ell) \in [118, 130]$~\gev. Stacked histograms follow the same color coding and normalisation as Fig.~\ref{fig:m4l}. The gluon-initiated and $Z/\gamma^* + \mathrm{jets}$ background distributions are scaled ($5\times$) for visualisation purposes.}\label{fig:mZ1}
\end{figure}

The selection is designed to extract signal candidates from events with no jet activity, $\slashed{E}_T < 25$~\gev, and exactly 2 pairs of OSSF leptons. The experimental treatment of preselected leptons is mimicked by requiring $\Delta R(\ell_i, \ell_j) > 0.02$, as well as $M_{\ell^+, \ell^{\prime-}} > 4$~\gev \, (irrespective of flavour), in order to suppress events with leptons originating from the decay of low-mass resonances. We further impose a cut on the leading lepton's $p_T > 20$~\gev, and require that at least two of the sub-leading leptons have $p_T > 10$~\gev. Pairs of OSSF leptons are combined into $Z_1Z_2$ candidates, where $Z_1$ corresponds to the $Z$ candidate with an invariant mass closest to the nominal $Z$-boson mass (91.1876~\gev)~\cite{PhysRevD.98.030001}, and $Z_2$ is the remaining one. Low-mass dilepton resonances produced along with an on-shell $Z$-boson are rejected by requiring $M(Z_1) \in [40, 120]$~\gev, as well as $M(Z_2) \in [12, 120]$~\gev. Finally, the mass range that characterises our signal region is defined as $M(4\ell) \in [118, 130]$~\gev, as shown in Fig.~\ref{fig:m4l}, which results in a $S/B_{\mathrm{irr}}$ ratio of 1.37. Upon including the yield of reducible-background events into account, the signal-to-background ratio $S/B$ gets reduced to 1.09. The effect of each cut on the fraction of retained events of the SM signal and irreducible backgrounds is shown in Table \ref{table:cutflow}, and the invariant mass distributions of our surviving $Z_i$ candidates are shown in Fig.~\ref{fig:mZ1}.

\subsection{Angular extraction}
\label{sec:subsec51}

\begin{figure}
    \centering
    \includegraphics[width=1.0 \textwidth]{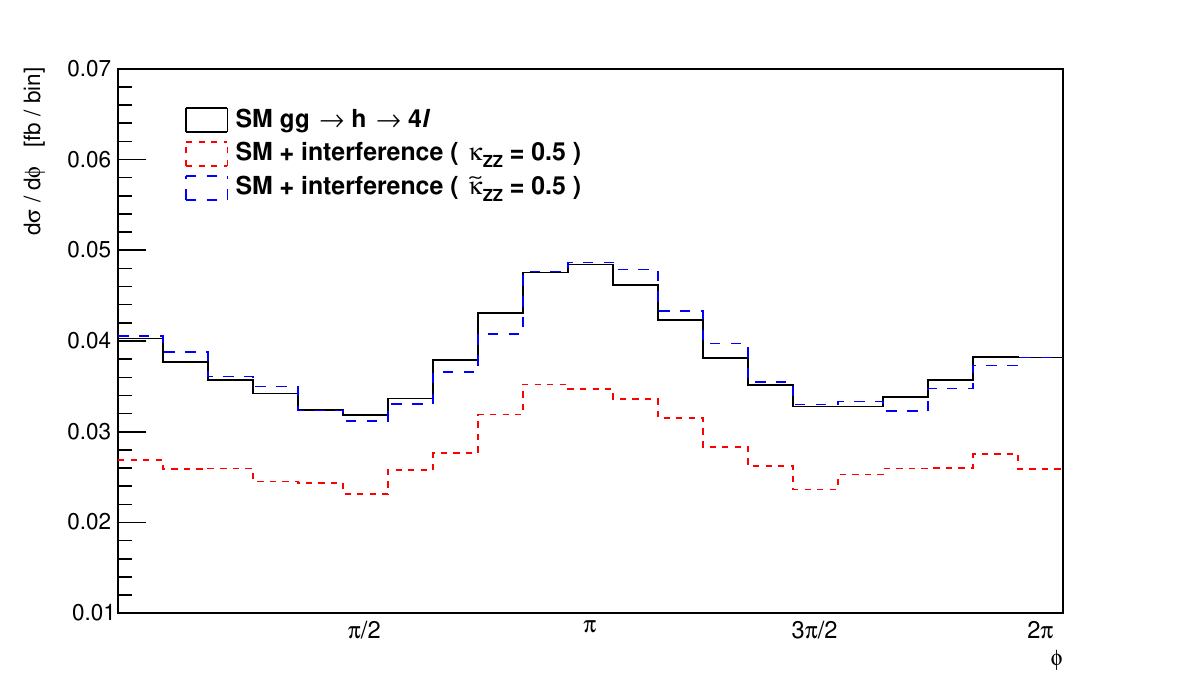}
    \caption{Azimuthal angle $\phi_{S'}$ differential distribution of the negatively-charged lepton (defined in the Higgs rest frame $S'$) for the SM-driven $gg \to h \to 4\ell$ process (solid black), as well as the SM~+~interference terms for the $CP$-even $\kappa_{ZZ} = 0.5$ (dotted red) and $CP$-odd $\tilde{\kappa}_{ZZ} = 0.5$ (dashed blue) operators.}
    \label{fig:phidist}
\end{figure}

Given the experimental limitations to determine the helicity of the final-state leptons, in what follows we restrict ourselves to define the various scattering angles with respect to the lepton with negative charge coming from the decay of the parent $Z_i$ boson ($i = 1, 2$), and which we refer to as $\ell^-_i$ as depicted in Fig.~\ref{fig:scatt_angles}. In order to implement the analysis of the differential distributions with respect to the angular variables described in Secs.~\ref{sec:sec3}-\ref{sec:sec4}, we Lorentz-boost the 4-lepton system to the centre-of-momentum frame $S'$, where the Higgs boson is at rest, and we have back-to-back $Z$-boson momenta. In frame $S'$, we construct a Cartesian coordinate system \{$\hat{x}$, $\hat{y}$, $\hat{z}$\} as follows: the $\hat{z}$ axis points in the direction of motion of $Z_1$; $\hat{y}$ is normal to the plane generated by $\hat{z}$ and $\hat{B}$, where $\hat{B} = (0, 0, 1)$ corresponds to the unit vector defining the beam direction in the laboratory frame; finally, $\hat{x}$ completes the right-handed set. Hence, we calculate the azimuthal angle $\varphi_{i,S'}$ formed in frame $S'$ between $\ell^-_i$ and the scattering plane as 
\begin{equation*}
    \tan \varphi_{i,S'} = \left(\frac{\ell^-_{i,\hat{y}}}{\ell^-_{i,\hat{x}}}\right),
\end{equation*}
where $\ell^-_{i,\hat{e}}$ corresponds to the projection of $\ell^-_i$ onto the $\hat{e}$ axis in frame $S'$, and $\varphi~\in~[0,2\pi)$. The azimuthal angle $\phi_{S'}$ between the planes formed by the lepton pairs in frame $S'$, used in our angular analysis, is then defined as
\begin{equation}
    \label{eq:phi}
    \tan \phi_{S'} \equiv \tan (\varphi_{2,S'} - \varphi_{1,S'}),
\end{equation}
with $\phi~\in~[0,2\pi)$. The azimuthal distribution for the process $gg \to h \to 4\ell$ is depicted in Fig.~\ref{fig:phidist}. Finally, each pair of OSSF leptons is further boosted to the rest frame $S''$ of its parent $Z$-boson, where we have back-to-back leptons momenta. The polar angle $\theta_{i,S''}$ between $\ell^-_i$ (in frame $S''$) and the direction of motion of its parent $Z_i$ boson~\footnote{The definition of the aforementioned angles is taken with respect to the SM-driven $gg \to h \to 4\ell$ process, where both pairs of leptons are produced from the decay of a $Z$ boson. However, it is important to note that a $g_{Zf}^h$ insertion, where a pair of leptons has no parent $Z$ boson, is also possible, and hence the direction of motion of the dilepton system in the Higgs rest frame $S'$ needs to be taken into account.} (in frame $S'$) is defined as
\begin{equation}
    \label{eq:theta}
    \cos \theta_{i,S''} \equiv \frac{\vec{p}_{\ell_i^-,S''} \cdot \vec{q}_{Z_i,S'}}{\left|\vec{p}_{\ell_i^-,S''}\right| \left|\vec{q}_{Z_i,S'}\right|},
\end{equation}
where $\vec{k}_{M_i,N}$ corresponds to the 3-momentum of particle~$M_i$ in frame $N$, and $\theta~\in~[0,\pi]$.

\section{Results}
\label{sec:sec7}

\begin{figure*}[!t]
\begin{center}
  \includegraphics[width=0.7\textwidth]{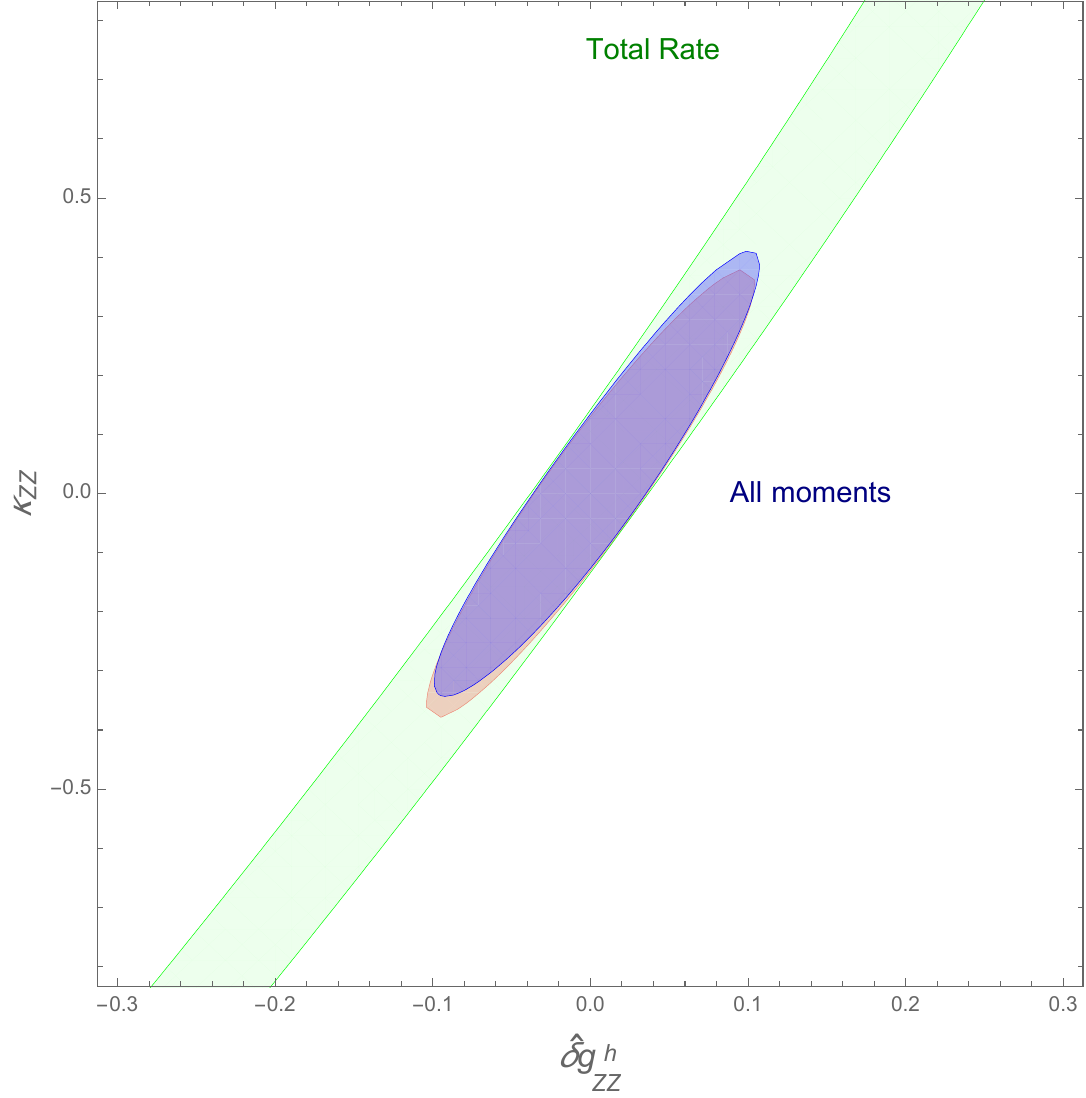} 
 \caption{Bounds at 68$\%$ CL on the $CP$-even anomalous couplings. The green band shows the bound from the total rate which keeps a flat direction, $\kappa_{ZZ}\approx 3.7~\delta \hat{g}^h_{ZZ}$ , unconstrained. The blue ellipse shows our final bounds including all the angular moments. The red ellipse also shows the results of an angular moment analysis but considering only the interference between the EFT and SM terms.}
  \label{bound1}
\end{center}
\end{figure*}

\begin{figure*}[!t]
\begin{center}
  \includegraphics[width=0.7\textwidth]{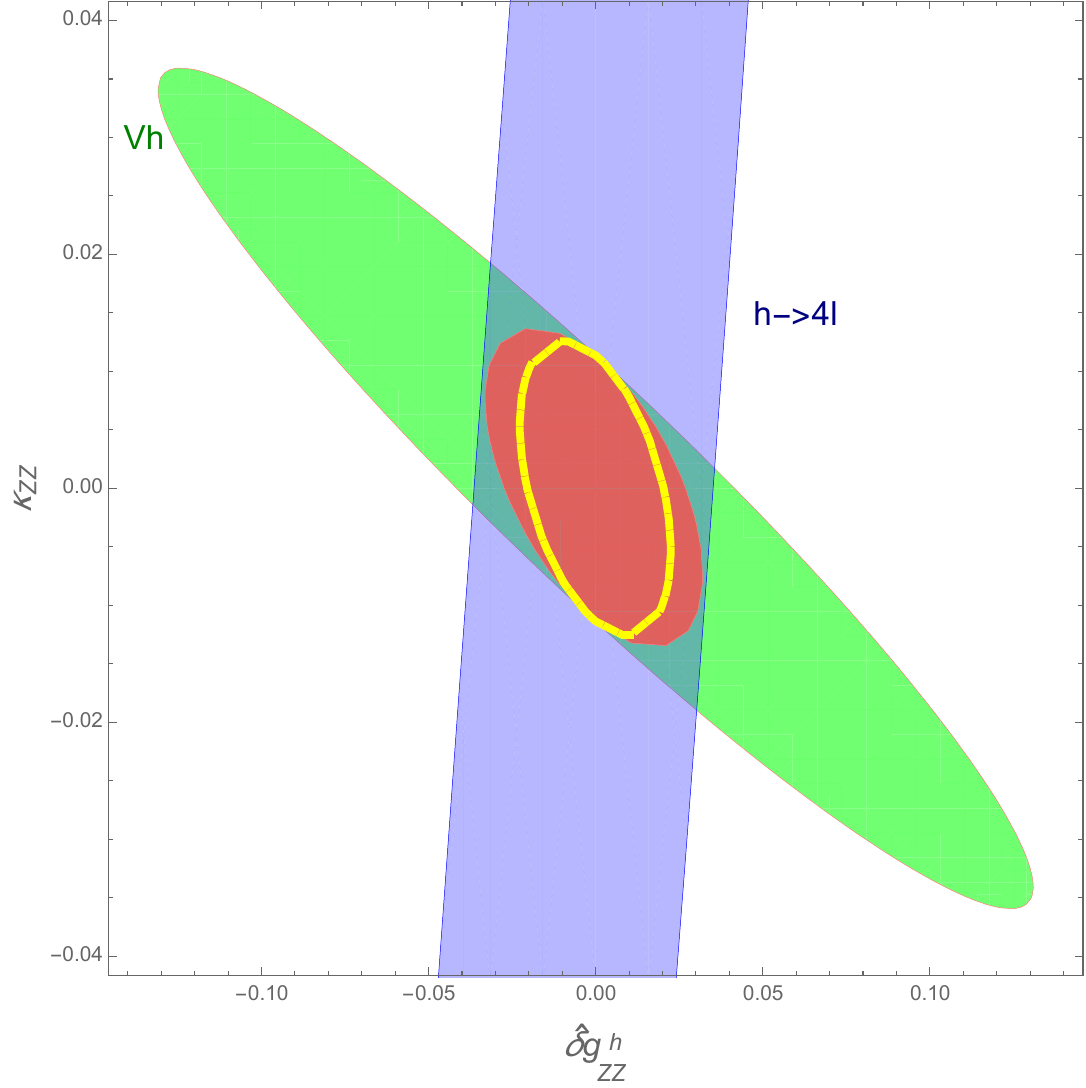} 
 \caption{Bounds at 68$\%$ CL on the $CP$-even anomalous couplings after combination with the results of the angular moment analysis of the $pp \to Wh/Zh$ processes carried out in Ref.~\cite{Banerjee:2019twi}. The blue band shows the results of this work, the green ellipse the bounds from $pp \to Wh/Zh$ processes. The red ellipse is the combination of the present work and moments analysis presented in Ref.~\cite{Banerjee:2019twi} for the $pp \to Vh$ processes. Finally, the dashed yellow ellipse shows the final bound after combination with  $pp \to h \to WW \to  2\ell 2 \nu$ process in Ref.~\cite{HLLHC}.}
  \label{bound2}
\end{center}
\end{figure*}

To obtain our bounds we define a $\chi^2$ function as follows,
\bea
\label{2chis}
\chi^2(\delta \hat{g}^h_{ZZ}, {\kappa}_{ZZ}, \tilde{\kappa}_{ZZ})&=& \sum_{ij} {\left(a_i^{EFT} - a^{SM}_i \right)}\Sigma^{-1}_{ij}{\left(a_j^{EFT} - a^{SM}_j\right)}\nn\\
\eea
where the covariance $\Sigma_{ij}$ is defined in Eq.~\ref{eq:ai_err}.  As explained in Sec.~\ref{sec:sec2} and~\ref{subsec:sec32} we do not include the $ g^h_{Zf}$ parameters of the $hZ\bar{\ell}\ell$ contact terms as they can be constrained very stringently using other processes. Using the above $\chi^2$ function, we obtain the 68$\%$ CL bounds shown in Fig.~\ref{bound1}. The green band shows the bound obtained if we include only information about the total rate of the process. It is clear that there is a flat direction, $\kappa_{ZZ}\approx 3.7~\delta \hat{g}^h_{ZZ}$, which can be constrained only by introducing the differential information of the angular moments. We also show in red the bound obtained if we only include the  interference term between the SM and EFT which almost coincides with the bounds that include the EFT squared term in blue. This implies that, if $\Lambda$ is the UV cutoff of the EFT, a truncation of the cross-section at the $1/\Lambda^2$ order includes already most of the BSM effects encoded by dimension-six operators; thus, there is no need of taking into account the $1/\Lambda^4$ level and an analysis without dimension-eight operators can be considered consistent.  If we assume no systematic uncertainties we obtain the bound $|{\kappa}_{ZZ}|< 0.05$ for  $\delta \hat{g}^h_{ZZ}=0$ which is comparable to the MELA bound $|{\kappa}_{ZZ}|< 0.04$~\cite{MELA}.

To obtain the bounds in Fig.~\ref{bound1} we have taken, $\tilde{\kappa}_{ZZ}=0$. We have, however, checked that a non-zero $\tilde{\kappa}_{ZZ}$ hardly changes the plot, i.e. to a very good approximation we will  obtain the same bounds if $\tilde{\kappa}_{ZZ}$ is marginalised over. This is because there is hardly any contribution to $\chi^2$ from $a_7$-$a_9$, the only moments that contain  a term linear in $\tilde{\kappa}_{ZZ}$. The  moments  $a_1$-$a_6$ that give the dominant contribution, on the other hand, depend on the $\tilde{\kappa}^2_{ZZ}$ which results in the  $\chi^2$ function being only mildly dependent on $\tilde{\kappa}_{ZZ}$. For the same reason we obtain a very weak bound on $\tilde{\kappa}_{ZZ}$,
\begin{equation}
|\tilde{\kappa}_{ZZ}|\lesssim 0.5
\end{equation}
after marginalising over $\delta \hat{g}^h_{ZZ}$ and ${\kappa}_{ZZ}$. This is unfortunately not competitive with  the projection, $|\tilde{\kappa}_{ZZ}|< 0.05$  from the $pp \to Wh/Zh h(bb)$ processes  in Ref.~\cite{Banerjee:2019twi}.

We can combine the above bounds with the projections from the $pp \to Wh(bb)/Zh(bb)$ and $pp \to h\to WW\to 2 \ell 2 \nu$ processes in Ref.~\cite{Banerjee:2019twi} and Ref.~\cite{HLLHC}. In order to combine these different processes,  we need to utilise the stringent constraints on the $Zff$ couplings, assume that the function $f$ in Eq.~\ref{eq:Higgsprod} and EFT deformations that rescale the $h \to bb$ branching  ratio can be independently constrained in a global fit including all the relevant Higgs physics processes. We also need to use EFT correlations, derived in Ref.~\cite{Banerjee:2019twi}, between anomalous gauge-Higgs couplings involving the $W$-boson and those involving the $Z$ boson. We can then combine the results of the angular moment analysis of $pp \to Wh(bb)/Zh(bb)$ with leptonic decays of the $W/Z$ in Ref.~\cite{Banerjee:2019twi} and the bound on the total rate for $pp \to h \to WW \to  2\ell 2 \nu$ in Ref.~\cite{HLLHC}.\footnote{An analysis using the method of moments is more challenging for the process as it is not possible to fully reconstruct  the angular information for the process because of the two neutrinos.} The final results are shown in Fig.~\ref{bound2}. We see that the complementarity of the $h \to VV^*\to V\ell \ell$ and $pp \to Vh$ processes, \textit{i.e.}, the fact that these processes probe very different linear combination of $\delta \hat{g}^h_{ZZ}$ and ${\kappa}_{ZZ}$, results in strong percent level bounds on these couplings.

A possible criticism towards our approach is that it is based on leading order matrix elements and is not optimised to include the effects of parton shower, detector effects and selection cuts. Before concluding this section we want to compare our results with a BDT analysis which does not have this shortcoming. We perform a simple BDT analysis with three variables, \textit{i.e.}, $\theta_1, \theta_2$ and $\phi$. We choose the Gradient Boosted Decision Tree (BDTG) algorithm as it can properly handle negative-weight events which arise upon including NLO samples. As $\delta \hat{g}^h_{ZZ}$ only rescales the SM matrix element and leaves no differential signatures, we only vary $\kappa_{ZZ}$ to obtain the bound $|\kappa_{ZZ}| < 0.052$, with a zero systematic uncertainty hypothesis. This can be compared with our earlier derived bound, $|\kappa_{ZZ}|<0.051$. As we can see, the numbers are very similar and the nominal difference is perhaps owing to statistical reasons. Thus, this validation shows that the presence of the experimental effects mentioned above fortunately do not affect the sensitivity of our method for this particular final state.

\section{Conclusions}
\label{sec:sec8}

We have carried out a fully differential study of the golden Higgs decay channel $h\to 4\ell$. The leptonic final state can be accurately reconstructed to give a wealth of differential information. In the Higgs rest frame three angles completely determine the direction of the final state leptons. A thorough differential study of the resulting three-dimensional space is one of the main experimental tools to probe the  tensor structure of the Higgs coupling to gauge bosons which includes new contributions in the dimension-six SMEFT.

In this work we show that the full angular distribution can be written as a sum of a set of basis functions both in the SM as well as in the dimension-six SMEFT.  The coefficients of these functions, the so-called angular moments, therefore encapsulate the full angular information of the process.
We derive the  analytical expressions for these angular moments including dimension-six SMEFT deformations. We then use the method of moments to extract these angular moments from our  Monte Carlo sample, which is simulated and analysed using a strategy that closely follows the LHC experiments. 

We finally use the extracted angular moments in the SM and dimension-six SMEFT to obtain projections for bounds on all the relevant  gauge-Higgs coupling deformations parametrised by  $\delta \hat{g}^h_{ZZ}, \kappa_{ZZ}$ and $\tilde{\kappa}_{ZZ}$ in Eq.~\ref{anam}. Our final results in Fig.~\ref{bound1} show that the angular moment analysis is crucial in eliminating flat directions that arise if one takes into account only the total rate of the process in the SM and SMEFT. Finally we combine our results with those of Ref.~\cite{Banerjee:2019twi} where a similar analysis using angular moments was carried out for the $pp \to V h$ process as shown in Fig.~\ref{bound2}. This combination allows us to obtain the strongest reported bounds  on the anomalous gauge-Higgs couplings.

\begin{acknowledgments}
We thank Silvia Ferrario Ravasio, Shilpi Jain, Lorenzo Moneta, Karl Nordstr\"{o}m, Marek Sch\"{o}nherr, and Gurpreet Singh Chahal for several helpful exchanges and discussions. We also thank Pietro Baratella for his insightful comments on the draft. OOV acknowledges the Mexican National Council for Science and Technology (CONACYT) - grant number - 460869. EV has been partially funded by the Deutsche Forschungsgemeinschaft (DFG, German Research Foundation) under Germany's Excellence Strategy - EXC-2094 - 390783311, by the Collaborative Research Center SFB1258 and the BMBF grant 05H18WOCA1
and thanks the Munich Institute for Astro- and Particle Physics (MIAPP) for hospitality. SB acknowledges the grant received from the IPPP, where the major part of this work was done.
\end{acknowledgments}

\bibliography{Biblio}

\end{document}